\documentclass[preprint]{aastex}

\usepackage{natbib,amssymb}
\newcommand{\beq}{\begin{equation}}
\newcommand{\eeq}{\end{equation}}
\newcommand{\beqa}{\begin{eqnarray}}
\newcommand{\eeqa}{\end{eqnarray}}
\newcommand{\avg}[1]{\ensuremath{\langle #1 \rangle}}
\newcommand{\sci}[2]{\ensuremath{#1 \times 10^{#2}}}

\begin{document}

\title{Transport of Ionizing Radiation in Terrestrial-like Exoplanet Atmospheres}

\author{David S. Smith\altaffilmark{1}, John Scalo, and J. Craig Wheeler\altaffilmark{2}}
\affil{Department of Astronomy, The University of Texas at Austin,
Austin, TX 78712} 

\altaffiltext{1}{Harrington Doctoral Fellow, NSF Graduate Research Fellow}
\altaffiltext{2}{E-mail: \texttt{\{dss,parrot,wheel\}@astro.as.utexas.edu}}

\begin{abstract}

The propagation of ionizing radiation through model atmospheres of
terrestrial-like exoplanets is studied for a large range of column
densities and incident photon energies using a Monte Carlo code we
have developed to treat Compton scattering and photoabsorption.
Incident spectra from parent star flares, supernovae, and gamma-ray
bursts are modeled and compared to energetic particles in importance.
Large irradiation events with fluences of 10$^6$--10$^9$ erg cm$^{-2}$
at the conventional habitable zone can occur at a rate from many
per day (flares from young low-mass parent stars) to $\sim 100$ per
Gyr (supernovae and gamma-ray bursts).  We find that terrestrial-like
exoplanets with atmospheres thinner than about 100 g cm$^{-2}$ block
nearly all X-rays, but transmit and reprocess a significant fraction
of incident $\gamma$-rays, producing a characteristic, flat surficial
spectrum.  Thick atmospheres ($\gtrsim 100$ g cm$^{-2}$) efficiently
block even $\gamma$-rays, but nearly all the incident energy is
redistributed into diffuse UV and visible aurora-like emission,
increasing the effective atmospheric transmission by many orders
of magnitude.  Depending on the presence of molecular UV absorbers
and atmospheric thickness, up to 10\% of the incident energy can
reach the surface as UV reemission.  For the Earth, between
$\sci{2}{-3}$ and \sci{4}{-2} of the incident flux reaches the
ground in the biologically effective 200--320 nm range, depending
on O$_2$/O$_3$ shielding.  For atmospheres thicker than $\sim 50$
g cm$^{-2}$ in the case of pure Rayleigh scattering and $\sim 100$
g cm$^{-2}$ in the case of O$_2$/O$_3$ absorption, the UV reemission
exceeds the surficial transmitted ionizing radiation. We also discuss
the effects of angle of incidence and derive a modified two-stream
approximation solution for the UV transfer. Finally, we suggest
that transient atmospheric ionization layers can be frequently
created at altitudes lower than the equilibrium layers that result
from steady irradiation and winds from the parent star.  We suggest
that these events can produce frequent fluctuations in atmospheric
ionization levels and surficial UV fluxes on terrestrial-like
planets.

\end{abstract}


\section{Introduction}

Planets orbiting the Sun and other stars are occasionally subjected
to large ionizing fluxes from astronomical sources, such as
gamma-ray bursts, supernovae, and flares from the parent star.
During these highly stochastic events, $\gamma$-ray, X-ray, and UV
irradiation affects planetary atmospheric chemistry through ionization
and heating, and biological activity through direct mutational
enhancement or sterilization.

The frequency of some types of these events has been estimated by
\citet{scaloWheeler02} and \citet{scalo03}; see also \S\ref{sec:srcs}.
Rates and fluences for intense parent star flares are much more
frequent for planets in the conventional habitable zone (continuous
liquid water, see \citealt{kasting93}) of low mass stars; this is
discussed briefly below in \S\ref{sec:srcs_flares}. Whatever the
radiation source, the significance of the above phenomena depends
strongly on the transparency of the atmosphere to the high-energy
radiation. For this reason, we have studied the propagation of
ionizing radiation (X-rays and $\gamma$-rays in our case) through
a suite of model terrestrial-like exoplanet atmospheres of various
column densities subjected to irradiation by various incident
spectra. We also considered effects due to the angle of incidence.

Previous work has considered atmospheric irradiation by specific
X-ray and $\gamma$-ray events (\citealt{brown73},
\citealt{kasturirangan76}, \citealt{omongainBaird76}, \citealt{gehrels03},
and references therein), but only in a terrestrial context and,
except for Gehrels et al.~who studied the O$_3$ chemistry, were
aimed at estimating only the altitude-dependent ionization. Some
treated the radiative transfer in full detail while others used a
simple exponential attenuation approximation. No past work has
considered the electron excitation of atomic and molecular lines
as a channel for redistribution of the ionizing flux to the
biologically important ultraviolet spectral range, a major theme
of the present paper.  Our work also explores a large range of
atmospheric column densities, follows the energy transfer in detail
using a Monte Carlo approach, including accurate Compton and
photoabsorption cross sections, and is generally concerned more
with events that may affect planetary life.  In particular, we are
interested in events strong enough to result in biologically
significant doses of radiation at the ground, as well as observable
atmospheric chemical effects (e.g., photolysis, ionization, and
heating).

In the present paper, we do not attempt to couple the radiative
transfer to the atmospheric chemical or thermal structure; instead
we assume an isothermal exponential atmosphere of given composition,
column density, and scale height.  This is an excellent approximation
for the high-energy radiative transfer, in which the energies
considered are much greater than typical molecular electronic binding
energies, so the radiative transfer is basically independent of the
chemical composition or thermal structure.  We postpone a calculation
of the impact on the chemical and thermal structure of the atmospheres,
since our main focus is on the radiation that reaches the planetary
surface.  Specific application of the results of this work for Earth
and Mars is given in \citet{smith04oleb}.  Additionally, we are
concerned with photon irradiation only. Energetic particles
from the parent star or the Galaxy would induce similar
processes and their energetics are briefly summarized in
\S\ref{sec:particles}.

Note that recent studies of irradiated ``hot Jupiters'' (e.g.,
\citealt{seagerSasselov98}) are concerned with the effect of the
mostly visual radiation of the central star on giant planet
atmospheres, not irradiation of terrestrial-like atmospheres by
high-energy radiation.  The present work has more in common with
studies of irradiation of accretion disk atmospheres by compact
high-energy radiation sources \citep{ross79,kallmanMccray82,rossFabian93}.
In addition, we confine our study to ionizing photons and neglect
radiation produced through high-energy particle cascades (see
\citealt{molinacuberos01} and references therein for the case of
Mars).

The kinds of events we are considering are of interest for their
effects on the stability of planetary chemistry, but our primary
motivation is the question of how and whether evolution would proceed
on habitable planets immersed in a highly variable radiation
environment.  This question certainly includes the Earth, since it
is likely that the terrestrial radiation environment has varied
significantly over timescales from millennia to eons due to a variety
of phenomena.  Would an intermittently enhanced and stochastic
radiation field and mutation rate inhibit or sterilize life, or
instead accelerate its occurrence and evolution?  In the simplest
theoretical models for evolution at the level of allele frequencies,
the rate of evolution is proportional to the variance of the allele
probabilities, and this variance is increased by mutations.  Thus
one might reason that hypermutation events greatly widens the degree
of variability on which selection can operate.  The genetic response
to hypermutation events or sudden environmental change, and the
rate of evolution of the mutation rate itself, is complex (see for
example \citealt{sniegowski00}).  There is already some experimental
evidence that evolutionary rates can be increased by enhanced rates
of mutation \citep{itoh02}.

We are envisioning planetary biospheres in which environmental
novelty and exogenous mutational variation is the norm.  There are
some tantalizing lines of evidence and arguments suggesting  that
diversity and hence evolution might be enhanced by an environment
that is complex.  Directed \emph{in vitro} and artificial life
evolution experiments both indicate that genome lengths (one metric
of complexity) grow only in information-rich environments (see
\citealt{adami00}). That the rate of evolution increases with
environmental diversity or novelty has been demonstrated in organisms
as simple as the yeast \emph{Pseudomonas fluorescens}
\citep{raineyTravisano98} and as complex as guppies \citep{reznick97}.
See also \citet{moxon94,peak96,vanbelkum98,pinaud02}, and
\citet{chessonHuntly97}.  The recognition of the existence of
hypermutability mediated by heat shock proteins (see
\citealt{rutherfordLindquist98}; \citealt{federHofmann99}, and the
overview by \citealt{pugliucci02}) and mutator mutases (e.g.,
\citealt{giraud01,radman99, radman00}) are of particular relevance
to the present work. Another line of evidence comes from considering
evolution as a learning process.  Experiments using neural networks
as the phenotype for digital genomes show that learning is more
efficient in the presence of bursts of strong mutation compared to
a constant mutation rate \citep{moriartyMiikkulainen95,
moriartyMiikkulainen99, gomezMiikkulainen97}. From one perspective,
mutations that increase fitness can be regarded as random measurements
on the environment, and genomes as selection-imprinted genetic
memory of past environments \citep{adami00}. Flares, supernovae,
and other stochastic radiation events may provide a wide information
channel on which natural selection can operate.

This point of view should be contrasted with the assumption of
\citet{wardBrownlee00} and \citet{gonzalez01} that development of
complex organisms primarily requires self-regulated stability of
the environment. Surely stability at some level (especially with
respect to temperature) is desirable for the continuance and evolution
of life, but between the extreme limits of tolerance, it is possible
that development of complexity is enhanced by fluctuations, as the
above examples suggest. If so, accelerated evolution and development
of complexity may occur on habitable planets subjected to a strongly
fluctuating radiation environment.  This possibility is obviously
of crucial importance for selection of targets in SETI searches
\citep{turnbullTarter03}.  While we cannot yet demonstrate whether
this occurs, or whether such an environment simply retards evolution
or sterilizes life, the arguments given above suggest that an
acceleration of evolution is plausible, and in any case motivates
us to quantify the nature of the fluctuations themselves, in
particular how well atmospheres of various column densities on
terrestrial-like exoplanets are buffered from jolts of ionizing
radiation.

\section{Methods}\label{sec:methods}

\subsection{Input}

\subsubsection{Exoplanet atmosphere model}\label{sec:atm_model}

Our work assumes an isothermal, plane-parallel atmosphere having
an exponential vertical density distribution.  Because of the high
energies of the photons compared to the molecular thermal energies
in the atmosphere, the value of the temperature is irrelevant for
our calculations except insofar as it affects the scale height.
Similarly, the photon energies are so large compared to electronic
binding energies of atoms or molecules that the radiative transfer
and secondary electron energy budget is approximately independent
of the composition of the atmosphere; in effect the photons only
``see'' free electrons at these energies. The particular density
values at each grid point are determined by specifying the total
column density in g cm$^{-2}$, which determines the number density
of particles at the ground $n_0$. Using $n_0$ and $h$, we build up
the grid according to $n_i = n_0 \exp(-z_i/h)$, where $z_i$ is the
altitude of the $i$-th grid point.

Choosing the scale height is fairly straightforward when using an
exponential density distribution. In all quantities related to the
radiative transfer, the scale height only appears in the product
$n_0 h$ and the quotient $z/h$.  The total column density of the
atmosphere is $n_0 h$, so fixing the column density (and hence the
optical depth) and adjusting the scale height only changes the
number density of molecules at the ground, which is irrelevant for
our work.  The one effect of choosing a particular scale height is
that it sets the altitude scale in terms of $z/h$, i.e., the
altitudes corresponding to given optical depths from the top of
the atmosphere are determined by $h$.  In this work, we use the
terrestrial scale height value of 8 km. The altitudes in our
results for any other desired scale height may be determined
according to \beq z' = \left(\frac{h}{\mathrm{8\ km}}\right)
z,\eeq where $z$ is the altitude given by our results, and $z'$ is
the altitude on the exoplanet with a scale height $h$.

If the habitability of an exoplanet is adequately defined by the
presence of liquid water, the lower limit to the column density of
a planet with surface liquid water is given by considerations of
water vapor photodissociation followed by atmospheric escape
\citep{ingersoll69, kasting88, kasting93}.  An estimate of the lower
limit is 30--100 g cm$^{-2}$ for a terrestrial mass planet, depending
on the temperature (J.~Kasting, private communication).  This is
also coincident with the smallest column density required to prevent
atmospheric collapse on synchronously rotating planets ($\gtrsim$
30 g cm$^{-2}$, as given by \citealt{joshi97}).

Avoiding speculation about the origins or robustness of life, for
illustrative purposes we assume that planets possessing even thin
atmospheres are indeed habitable---given enough warming flux from
the parent star.

Selecting atmospheric constituents for terrestrial-like exoplanets
is extremely speculative. Fortunately, we can neglect atmospheric
composition effects on the radiative transfer of the incident
ionizing radiation because the energies considered here are so
much larger than any internal atomic or molecular transition
energies of interest. The composition matters only in that it
determines the column abundances of electrons, which are the
primary scatterers. For simplicity, we chose an inert N$_2$
atmosphere for the transfer of ionizing radiation. None of the
results for the ionizing radiation would change significantly if
the primary constituent were CO$_2$ or O$_2$.

The transfer of the UV reemission \emph{does} depend on assumed
composition (see \S\ref{subsec:uvredist}). For this calculation,
we chose two simple UV opacity sources that probably bracket the
extremes of UV transparency. For present-day Earth analogues, we
included O$_2$ and O$_3$ absorption, distributed in an abundance
profile similar to Earth's (taken from Appendix C of \citealt{brasseur99}).
For column densities other than Earth's, we scaled the terrestrial
O$_3$ altitude profile to match the vertical O$_3$ column density
from the top of the exoplanet atmosphere, i.e., the peak of the
O$_3$ profile is always at the same column density from the top.
This could be a severe approximation---see \S\ref{sec:screening}.
For Archean Earth analogues, we removed molecular absorption and
included only Rayleigh scattering, although we recognize the
possibility of other Archean UV screens (e.g., \citealt{saganChyba97}),
especially aerosols.  These two idealizations are useful to gauge
the surficial UV fluence in different limits.

\subsubsection{Incident radiation spectra}

Given that we are interested in such sources as supernovae,
gamma-ray bursts, and stellar flares, we chose incident
spectra characteristic of these.  For supernovae we assumed
monoenergetic spectra of energies 0.125, 0.25, 0.5, 1, and 2 MeV
(1 MeV = \sci{1.24}{-3} nm), which correspond to the energy ranges
of $^{56}$Co and $^{56}$Ni decay lines \citep{hoeflich98}. We chose
not to model specific supernovae spectra---they are merely the
motivation for using a monoenergetic spectrum, which is instructive
in its simplicity. In order to model the effects of stellar flares
and gamma-ray bursts, we included two types of continuous spectra.

To represent the lower-energy spectra of flares, we adopted the
following parameterized formula which is actually the energy
dependence expected  for a high-temperature thermal plasma, such
as the solar corona \citep{tuckerKoren71}: \beq\label{eq:tuckerKoren}
\frac{dN}{d\lambda}\propto\lambda^{-2}\exp(-\lambda_p/\lambda),\eeq
where $dN/d\lambda$ is the number of photons per unit wavelength.
Our adopted spectral form for flares is also consistent with the
0.9--10 keV spectrum of the giant X-ray flare in the dMe star EV
Lac presented by \citet{favata00}. We are unaware of the expected
or observed form of generic solar or stellar flare spectra; the
problem is complicated by the fact that flares peak in different
wavelength regimes at different times. For the energy spectrum
corresponding to Eq.~\ref{eq:tuckerKoren}, the photon number
distribution can be shown to be \beq \frac{dN}{dE}= \frac{N_\gamma}{E_p}
\exp\left(-\frac{E}{E_p}\right), \eeq where $N_\gamma$ is the total
number of photons in the model and $E_p=hc/\lambda_p$ is the energy
corresponding to the peak of the wavelength spectrum
(Eq.~\ref{eq:tuckerKoren}).  We calculated models irradiated by
flare spectra with peak energies, $E_p$, of 2.2, 22, and  220 keV,
corresponding to average energies of 1, 10, and 100 keV (see
\citealt{kruckerLin02}. For these spectra, the lower and upper
photon energy cutoffs are $0.01 E_\mathrm{inc}$ and $4 E_\mathrm{inc}$,
respectively, where $E_\mathrm{inc}$ is the specified average
incident energy, from which the peak energy, $E_p$, was calculated.
We chose to place the above arbitrary limits on the flare spectra
for computational reasons: photons of higher energy than about $4
E_\mathrm{inc}$ are too improbable, and photons of energy much lower
than $0.01 E_\mathrm{inc}$ are of little physical significance
because of the low energies.

For the gamma-ray burst spectra, we used a broken power law with an
exponential cutoff known as a Band spectrum \citep{band93}. The
photon number distribution is given by \beq \label{bandeqn1}
\frac{dN}{dE}= k\left(\frac{E}{\mathrm{100\ keV}}\right)^\alpha
\exp(-E/E_0) \eeq for $E \leq (\alpha-\beta) E_0$ and \beq
\label{bandeqn2} \frac{dN}{dE}= k\left[
\frac{(\alpha-\beta)E_0}{\mathrm{100\ keV}} \right]^{\alpha-\beta}
e^{\beta-\alpha} \left(\frac{E}{\mathrm{100\ keV}}\right)^\beta
\eeq for $E \geq (\alpha-\beta)E_0,$ where $E_0 = 250$ keV is the
turnover energy, $\alpha$ and $\beta$ are the power law indices.
We adopt for this particular spectrum upper and lower wavelength
limits of 50 keV and 3 MeV, respectively, with an average energy
of 200 keV (constrained by observations, e.g.,
\citealt{preece00}).  The empirical power law indices lie in the
range $-1.6 \lesssim \alpha \lesssim 0.0$ and $-4.5 \lesssim \beta
\lesssim -1.5$ \citep{tavani00}. For our calculations, we adopt
$\alpha=-0.9$ and $\beta=-2.3$, which are roughly the averages in
the histograms given by Tavani et al.

We examined Band spectra, but for simplicity our results are given
only for the model flare spectra and monoenergetic spectra. Since
both the model flare spectra and Band spectra decline in photon
number at higher energies, the results are very similar if the
average energies are equal. Indeed, some of our results will be
shown to be completely independent of the form of the incident
spectrum.  Additionally, gamma-ray bursts are such infrequent
events that their contribution to the mutational environment of an
exoplanet is much smaller than supernovae and stellar flares.

The angle of incidence was taken to be normal to the atmospheric
boundary surface for most of the calculations.  Given a point
source in the sky, the angle of incidence will be roughly
perpendicular for most of the planet.  But the spherical symmetry
of the problem still makes a calculation for normal incidence
slightly more optimistic than one that included the full radiative
transfer effects of varying the angle of incidence. We tested
varying the angle of incidence and its effects on the transmitted
fraction and discuss the results briefly in \S\ref{sec:angInc}
below.

\subsection{Incident ionizing radiation transfer}

The transfer of the incident X-rays and $\gamma$-rays was handled
via a Monte Carlo code that was written for this work and that
accurately accounts for the complicated angular and energy dependences
of the cross sections. Appendix \ref{sec:mcalg} explains the algorithm
in detail.

The initial step of the calculation involves Compton scattering and
photoabsorption of $\gamma$-rays and X-rays.  Compton scattering
was implemented as an inelastic scattering cross section given by
the Klein-Nishina formula \citep{lingenfelterRothschild00}.
Photoabsorption was included as a purely absorptive cross section
of the empirical form \beq \label{eqn:pa} \sigma_{\rm pa}(E,Z) =
2.04\times 10^{-30} (1+0.008 Z) \frac{Z^3}{E^3}\ \mathrm{cm}^2,
\eeq where $Z$ is the atomic number of the absorber and $E$ is the
photon energy in units of the electron rest mass \citep{setlowPollard62}.
We found that Eq.~\ref{eqn:pa} reasonably represented the detailed
cross section measurements \citep{henke93} for a variety of elements
for energies greater than the corresponding K photoabsorption edge
(480 eV for nitrogen).

Since terrestrial-like exoplanet atmospheres can be very optically
thick to high-energy radiation (e.g., the optical depth is 65 at 1
MeV on the Earth), two weighting procedures were used to more
efficiently track photon statistics (see \citealt{watsonHenney01}
for a summary of weighting and other variance reduction techniques).
In our model, approximately 10$^6$ photons are initialized at the
top of the atmosphere heading downward with energies sampled according
to the specified incident spectrum.  Supernovae and flare spectra
were sampled by inversion, while the Band spectrum was sampled by
a rejection technique \citep{hammersleyHandscomb79,kalosWhitlock86}.
Every photon carries a statistical weight, which signifies the
probability that it is still scattering in the atmosphere after
each interaction.  Each photon is propagated to a random optical
depth, sampled from $e^{-\tau}$, since the probability that the
photon will travel a distance corresponding to $\tau$ without
interaction is $e^{-\tau}$. The photon is then statistically forced
to scatter by subtracting a fraction of its weight equal to the
probability, $e^{-\tau}$, that it did not scatter before exiting
the grid. This technique is known as ``forced scattering''
\citep{witt77}. In this way, the statistics are more accurately
tracked for a discrete number of photons, and each Monte Carlo
interaction explores many possible outcomes. Mersenne Twister
\citep{matsumotoNishimura98} was used for pseudo-random number
generation.

During the above process, we tracked the spectra of photon energy
deposited at the ground, electron energy deposited (via Compton
recoil and photoabsorption) at various layers in the atmosphere,
and photons lost to space. Later this information is used to calculate
the UV reemission fluxes and ionization fractions.

\subsection{UV redistribution}\label{subsec:uvredist}

\subsubsection{Physical process}

The Earth receives a steady flux of solar wind ions with very high
kinetic energies.  Energetic electrons produced via a variety of
mechanisms (fast particles, magnetohydrodynamic flows, etc.) excite
atmospheric constituents, resulting  in dynamic auroral displays
that extend from the ultraviolet to the infrared.  \citet{chamberlain61}
discusses the detailed physical mechanisms.

When astrophysical bursts of radiation, such as stellar flares,
supernovae, and gamma-ray bursts irradiate a terrestrial-like exoplanet with a
sufficiently thick atmosphere, analogous phenomena will occur. The
initial ionizing radiation creates primary electrons as a photoproduct.
These very energetic charged particles then produce secondary
photoelectrons which excite molecules and create aurora-like emission
in much the same way the solar wind and EUV does on Earth.

Since the energies considered here are so high, an incident photon
can cause the ionization of tens of thousands of molecules before
being absorbed.  Primary Compton-recoil electrons and photoelectrons
are responsible for the ionization as they are slowed by collisions
with neutral N$_2$ molecules. Each of these ionizations results in
a secondary electron, and it is these liberated electrons that
dominate the particle flux from the incident radiation (see
\citealt{evans74} for direct observations of auroral electron
spectra).  The average energy of secondary electrons released by
primary ionization of N$_2$ is about 35 eV \citep{fano63}, with
only a very weak dependence on primary electron energy or charge
of the target. For example the average energy per ion pair for air,
argon, and water are 34, 26, and 30 eV, respectively (see, for
example, \citealt{fano63}). The distribution of secondary energies
has been studied experimentally by \citet{peterson71,peterson72}.

As each secondary electron moves through the atmosphere, it can
exchange energy with other particles by (i) elastic Coulomb
interactions and elastic collisions with neutrals (both of which
lead to thermalization of the electron energy) or (ii) excitation
of internal degrees of freedom in the target molecules.  In case
(ii), secondary electron impact excitation of electronic, vibrational,
and rotational levels will result in a rich line spectrum extending
from the UV (electronic transitions) to the radio (pure rotational
transitions). This excitation by secondary electrons and subsequent
line emission is equivalent to the main process giving rise to the
terrestrial auroral spectra.  In our case the process redistributes
some of the energy of the X-ray and $\gamma$-ray photons into UV
and longer wavelengths. This redistribution is also analogous to
that which occurs in gases of cosmic abundances in accretion disks
around compact stellar remnants \citep{ross79,kallmanMccray82,rossFabian93}
and in interstellar clouds where secondaries from cosmic-ray
ionization events can result in a rich UV line spectrum
\citep{prasadTarafdar83, gredel89}. An important difference between
the molecular and atomic cases is that in the atomic case the
secondary electron energy must be thermalized once its energy falls
below the excitation potential of the first excited state of the
atom, whereas the molecular case has a broad spectrum of excitation
channels at lower energies.

\subsubsection{Excitation dominates heating}

In a highly ionized plasma, most of the electron energy would be
thermalized by electron-electron collisions because of the long-range
nature of the Coulomb interaction---very little of the energy would
go into excitation and line radiation. The importance of electron-electron
collisions depends on the ionization fraction, however, and for a
nearly neutral planetary atmosphere, most of the secondary electron
energy goes into excitation, not heating.  \citet{foxVictor88}
presented detailed calculations of the dependence on the ionization
fraction of the number of excitations to various electronic levels
of N$_2$.

We can derive an order of magnitude condition for excitation to be
more important than Coulomb interactions by comparing the respective
collision frequencies. The characteristic Coulomb collision frequency
can be expressed as (\citealt{spitzer78}, Eqs.~4.13 and 4.14) \beq
\nu_\mathrm{ee}  = \frac{4\pi e^4 n_e}{m_e^2 w^3} \ln\left(\frac{\Lambda
m_ew^2}{3kT}\right),\eeq where $w$ is the relative velocity between
test and field electrons and the factor \beq\Lambda \equiv
\left(\frac{9k^3 T^3}{4\pi n_e e^6}\right)^{1/2}\eeq is the usual
approximate cutoff factor in the Coulomb logarithm---see \citet{spitzer62}
and \citet{mitchnerKruger73} for derivations. The extra ratio of
energies in the logarithm accounts for the fact that the test
particles follow a non-Maxwellian velocity distribution. Noting the
$w^{-3}\propto E^{-3/2}$ dependence, we see that this agrees very
well with the analytical fit to more detailed calculations given
by \citet{swartz71}: \beq\label{eq:eeColl} \nu_\mathrm{ee} =
\sci{2.0}{-4} n_e^{0.94} E^{-1.44} \mathrm{sec}^{-1}, \eeq where
$E$ is the energy of the secondary electrons in eV and $E\gg kT$.
We adopt this convenient fit here. We neglect electron-ion Coulomb
scattering because the time scale for thermalization by electron-ion
scattering is larger than for electron-electron scattering because
the high electron-ion mass ratio reduces the per-collision energy
transfer efficiency.

We estimate the inelastic collision frequency to be \beq \nu_\mathrm{inel}
= n\sigma_\mathrm{inel}(2E/m_e)^{1/2}.\eeq The cross sections for
ionization and excitation of N$_2$ and other atmospheric gases are
energy dependent, with much structure due to resonances with dominant
electronic and vibrational transitions as the secondary electron
energy decreases.  A useful plot of cross sections for N$_2$, O$_2$,
and O from 1 to 100 eV is given in \citet{banksKockarts73}, and
cross sections for N$_2$ at low energies are given in Fig.~II.43
of \citet{mitchnerKruger73}.  \citet{edgar73} give cross sections
for five ionization continua of N$_2$ due to electron impact, with
cross sections of 10$^{-16}$ to 10$^{-17}$ cm$^2$ at 100 eV. The cross
sections are smaller by about a factor of three at 35 eV and decline
rapidly at still lower energies. (The first ionization potential
of N$_2$ is 14.5 eV.) Similar behavior is expected for other candidate
dominant constituents of planetary atmospheres. For N$_2$, below
about 20--30 eV typical inelastic cross sections are of order
10$^{-16}$ cm$^2$ down to about 1.5 eV, with variations of a factor
of a few (e.g., the local peak at about 2 eV due to excitation of
vibrational levels within the ground electronic states).  Similar
cross sections occur for other candidate molecules and for the
thermal inelastic electron impact excitation of atoms inferred from
data in \citet{spitzer78}. Taking this value of 10$^{-16}$ cm$^{-2}$
for the cross section, we estimate the inelastic collision frequency
to be \beq\label{eq:inelColl} \nu_\mathrm{inel} = \sci{5}{-9} n
\sigma_\mathrm{inel,16} E^{1/2}\ \mathrm{sec}^{-1},\eeq where $E$
is in eV, $n$ is in cm$^{-3}$, $ \sigma_\mathrm{inel,16}$ is in
units of 10$^{-16}$ cm$^2$.  Comparing Eqs.~\ref{eq:eeColl} and
\ref{eq:inelColl} we find that inelastic excitation will dominate
Coulomb thermalization ($\nu_\mathrm{ee} \ll \nu_\mathrm{inel}$)
when \beq \label{eq:maxIonFrac} n_e/n \ll \sci{4}{-2}
\sigma_\mathrm{inel,16} E_{35}^2,\eeq where $E_{35}$ is in units
of 35 eV. According to \citet{crisp00}, the ionization fractions
in the D ($\sim 90$ km), E ($\sim 110$ km), F1 ($\sim 170$ km) and
F2 ($\sim 300$ km) layers of the Earth's ionosphere are only
10$^{-12}$, 10$^{-7}$, 10$^{-5}$, and 10$^{-3}$, respectively. As
our results will show, all but the most extreme cases of irradiation
(such as a 10$^8$ erg cm$^{-2}$ stellar flare) will produce ionization
fractions below the limit of Eq.~\ref{eq:maxIonFrac}. Thus \emph{the
secondary electrons will expend nearly all their energy in excitation
and almost none in heat.}

Another portion of the secondary electron energy will be expended
in elastic, electron-neutral, molecular collisions. The electric
field of the electron polarizes the charge distribution in the
molecule, inducing a dipole moment, leading to an effective potential
at large distance that varies as $r^{-4}$ and a cross section that
varies as $w^{-1}$ (recall $w$ is the relative velocity). The
calculated and measured momentum transfer cross sections for such
interactions are large, of order 10$^{-15}$ cm$^2$ at the energies
of interest (10 times larger than for inelastic collisions). Despite
the large cross section, the fractional energy lost by the secondary
electron in a typical electron-neutral collision is of order
$2m_e/Zm_p$ (e.g., \citealt{mitchnerKruger73} Eq.~7.5), which makes
this process much less than a 1\% effect compared to excitation,
and so we neglect it.

\subsubsection{Approximate treatment of UV reemission}

Although the secondary electrons have a distribution of energies,
their mean energy is 35 eV for N$_2$, a value which is known to be
nearly independent of the composition, as discussed above. The
electron excitation cross sections as a function of energy for
N$_{2}$ have broad maxima around 10--80 eV (\citealt{jones74},
Fig.~4.15), which neatly brackets the average energy of the
secondary electrons, so we expect the collisionally excitable
N$_2$ electronic states to be well populated among the the target
molecules.  This suggests that, in the UV, the sources of strong
reradiation will be N$_2$ emission bands, similar to the case for
auroral lines \citep{jones74}. Data for some of the more important
band systems are given in Table \ref{table:auroralLines}
\citep{banksKockarts73,lofthusKrupenie77,huberHerzberg79}. We
emphasize that we have chosen a pure N$_2$ atmosphere simply to
keep the calculations and presentation manageable, and that any
molecule which might be suspected to dominate the compositions of
terrestrial-like exoplanet atmospheres has similarly spaced
electronic levels and should be excited with comparable
efficiency.

\begin{table}
\begin{center}
\begin{tabular}{lcc}
\hline
Transition & Species & Wavelength range \\
\hline A$^3\Sigma_u^+$-X$^1\Sigma_g^+$ (Vegard-Kaplan) & N$_2$ &
210--540 nm \\
a$^1\Pi_g$-X$^1\Sigma_g^+$
(Lyman-Birge-Hopfield) & N$_2$   &  130--200 nm \\
E$^3\Sigma_g^+$-A$^3\Sigma_u^+$ (Herman-Kaplan) & N$_2$   &
213--274 nm \\
C$^3\Pi_u$-B$^3\Pi_g$           (2nd positive) &
N$_2$   & 268--545 nm \\
B$^2\Sigma_u^+$-X$^2\Sigma_g^+$ (1st
negative) & N$_2^+$ & 320--600 nm \\
\hline
\end{tabular}
\caption{\label{table:auroralLines}Strongest UV N$_2$ and N$_2^+$
electronic band systems
\citep{banksKockarts73,lofthusKrupenie77,huberHerzberg79}.}
\end{center}
\end{table}

We ignore the complication of the full line radiative transfer,
since we are interested in only estimating the transparency of
atmospheres to auroral emissions. The density and amplitude of lines
(in photon number per unit wavelength) in auroral spectra is roughly
distributed uniformly from the UV to the near IR (see spectra in
\citealt{jones74, chamberlain61}), so we assume the energy fluence
$F_\mathrm{dep,i}$ deposited at each layer $i$ is reradiated from
that layer in the form \beq \label{eqn:uvreemit}
\frac{dF_\mathrm{UV,i}}{d\lambda} = \frac{F_\mathrm{dep,i}}{\lambda
\ln(\lambda_\mathrm{max}/\lambda_\mathrm{min})} \eeq between the
wavelengths $\lambda_\mathrm{min}$ and $\lambda_\mathrm{max}$
corresponding to the lower and upper limits of the important auroral
emission lines. The Monte Carlo calculation yields the fraction of
the original incident energy that is deposited by X-ray photoabsorption
and Compton recoil at each layer $i$.  From this number, we assume
that all primary electron energy is transferred to secondary
electrons. At each layer a spectrum of the form of Eq.~\ref{eqn:uvreemit}
is reemitted isotropically and then attenuated either by Rayleigh
scattering or molecular absorption.

We have replaced the rich and extremely complex line spectrum of
N$_2$ (and other molecules) by a continuous spectrum that contains
(roughly) the same amount of flux per unit wavelength interval as
the line spectrum. This smearing of the line spectrum into an
equivalent continuous spectrum was assumed because: (i) we are
interested in only an order-of-magnitude estimate for the fraction
of energy that reaches the ground in each wavelength interval, and
(ii) the alternative would require the solution of a large number
of rate equations for the level populations at each altitude, a
calculation beyond the scope of the present work.

\subsubsection{Atmospheric UV screening}\label{sec:screening}

Any UV reemission will be subject to a variety of opacity sources
within the exoplanet atmosphere.  Depending on the precise atmospheric
composition, the primary UV screens might be molecular absorbers
or aerosols, or in the absence of these, pure Rayleigh scattering.
We take two extreme limits: pure O$_3$/O$_2$ absorption with a
terrestrial abundance profile (characteristic of present-day Earth)
and pure Rayleigh scattering (i.e., no molecular or aerosols
absorbers, characteristic of Archean Earth).  Although we recognize
that there is considerable uncertainty concerning UV screening in
the Archean atmosphere (e.g., \citealt{levyMiller98, cockell02}),
recent evidence concerning mass-independent isotopic fractionation
in Archean sulfides \citep{farquhar02} suggest the absence of a
significant UV shield during this period \citep{wiechert02}, so our
assumed Archean atmosphere may not be so extreme.  And if fluxes
are large enough to significantly erode the ozone layer (e.g.,
\citealt{gehrels03} and references therein), then the pure scattering
case may be relevant even for periods when the planet
possessed an ozone layer.

To find the fraction of the reemitted flux that would reach the
ground in the case of pure Rayleigh scattering, we attenuated the
reemission on a layer-by-layer basis according to a modification
of the \citet{schuster05} solution for ``foggy'' atmospheres,
which is a special case of the two-stream approximation (see
Appendix \ref{sec:schusterModification} for the full derivation).
In this scheme, the fraction of the flux emitted at layer $i$
transmitted by the atmosphere is \beq T(\lambda,z) = \frac{1/2 +
\tau_\uparrow(\lambda,z)}{1+\tau_\uparrow(\lambda,z)+\tau_\downarrow(\lambda,z)},
\label{eqn:tscatt}\eeq where $\tau_\downarrow$ and $\tau_\uparrow$
are the optical depths of the part of the atmosphere below and
above the layer of reemission, respectively. We assume, to good
approximation, that the emission layer itself is of negligible
optical depth (we use 256 altitude zones per atmosphere). The
redistributed UV flux received at the ground, $F_\mathrm{UV}$, is
then  \beq \label{eqn:uvflux}F_\mathrm{UV} =
\int_0^{z_\mathrm{max}}
\int_{\lambda_\mathrm{min}}^{\lambda_\mathrm{max}}
F_\mathrm{UV}(\lambda,z)\ T(\lambda,z)\,d\lambda\,dz,\eeq where
$F_\mathrm{UV}(\lambda,z)$ is the differential photon number
spectrum as a function of wavelength and altitude,
$z_\mathrm{max}$ is the altitude of the highest atmosphere zone,
and $T(\lambda,z)$ is the wavelength-dependent transmission
function for layer at height $z$ given by Eq.~\ref{eqn:tscatt} for
the optical depths above and below that layer. The surface
transmitted energy fractions we calculate in this manner are upper
limits, since we neglect aerosol absorption and scattering and
collisional deexcitations (see Appendix \ref{sec:quench}).

A very different situation occurs if the atmosphere contains a
significant source of UV molecular opacity at altitudes below the
bulk of the secondary electron deposition.  We use O$_2$ and O$_3$
as our prototype. To examine molecular absorption, we must neglect
Rayleigh scattering, since the above treatment applies only in the
pure scattering limit. We assume the transmission through each layer
in the presence of molecular absorbers follows the Beer-Lambert
law, so that \beq T(\lambda,z) = \exp[-\tau(\lambda,z)],\eeq where
$\tau(\lambda,z)$ is the wavelength-dependent optical depth to
absorption from height $z$ to the ground.  The subsequent calculation
of $F_\mathrm{UV}$ is analogous to Eq.~\ref{eqn:uvflux}. Integrating
over atmospheric layers is equivalent to a formal solution to the
transfer equation (neglecting the angular dependence), in which the
source function at each layer is due only to redistributed UV
radiation, since the thermal contribution at these wavelengths is
negligible for any possible atmospheric temperature. For this case,
we assume that half of the reemitted flux is directed straight
downward and half is directed upward. We ignore the upward fraction
and attenuate the downward half to obtain our estimate.

We chose terrestrial fractional abundances of O$_2$ and O$_3$ (taken
from \citealt{brasseur99}) for our absorption case.  Since the
relative ozone is to zeroth order a photoproduct of irradiation
incident on the top of the atmosphere, the ozone concentrations as
a function of column density from the top of the atmosphere should
be approximately invariant.  Taking ozone on the Earth as the
prototype, we scaled terrestrial concentrations to match our various
atmosphere models.  In each case, the fractional abundance of ozone
at a particular altitude on the exoplanet was matched with the
abundance of terrestrial ozone at an altitude corresponding to the
same column density from the top of the atmosphere.  In cases where
the exoplanet atmospheric column density was smaller than that of
Earth, we truncated the ozone profile at low altitudes.  It must
be noted that this is a gross simplification, since the O$_3$ profile
depends on the O$_2$ column density, not the total optical depth,
and even then the O$_3$ peak does not follow the O$_2$ column density
linearly because of the density-dependence of the ozone production
from O $+$ O$_2$ \citep{kastingDonahue80,kasting85}.  For this
reason the calculations of ozone shielding of the redistributed UV
radiation must be considered as illustrative only.  The
wavelength-dependent O$_2$ and O$_3$ cross sections were taken from
\citet{yungDemore99}. The transmitted fractions in the presence of
the ozone shield presented hereafter assume an upper limit to the
biologically relevant flux of 320 nm. Unfortunately, the results
are very sensitive to this quantity (as shown in Fig.~\ref{fig:uvCutoff}),
but evidence from terrestrial UV-B damage (see \S\ref{sec:uvthick})
supports this conservative value.

The effects of collisional deexcitation of target molecules is
addressed in detail in Appendix \ref{sec:quench}.  We find that, for the
strongest auroral nitrogen lines, this mode of energy dissipation
is unimportant.

\section{Results and discussion}
\label{sec:results}

\subsection{Habitable exoplanet surfaces can be exposed to significant 
$\gamma$-ray---but not X-ray---fluences}

According to our Monte Carlo calculation, exoplanet atmospheres
with column densities between roughly 30 and 100 g cm$^{-2}$
(habitable by our definition in \S\ref{sec:atm_model}) will transmit
at least 1\% of the incident $\gamma$-rays to the surface. Furthermore,
for atmospheres between 30 and 50 g cm$^{-2}$ and a source of MeV
photons such as a supernova or gamma-ray burst, most of the surficial energy
fluences will be due to the incident ionizing radiation.  The
characteristic energy of the radiation received at the ground on
planets with thin atmospheres will be very high since the redistribution
of the radiation through electron-mediated excitation processes
discussed above will be small due to the low optical depths.

\begin{figure}
\centerline{ \plotone{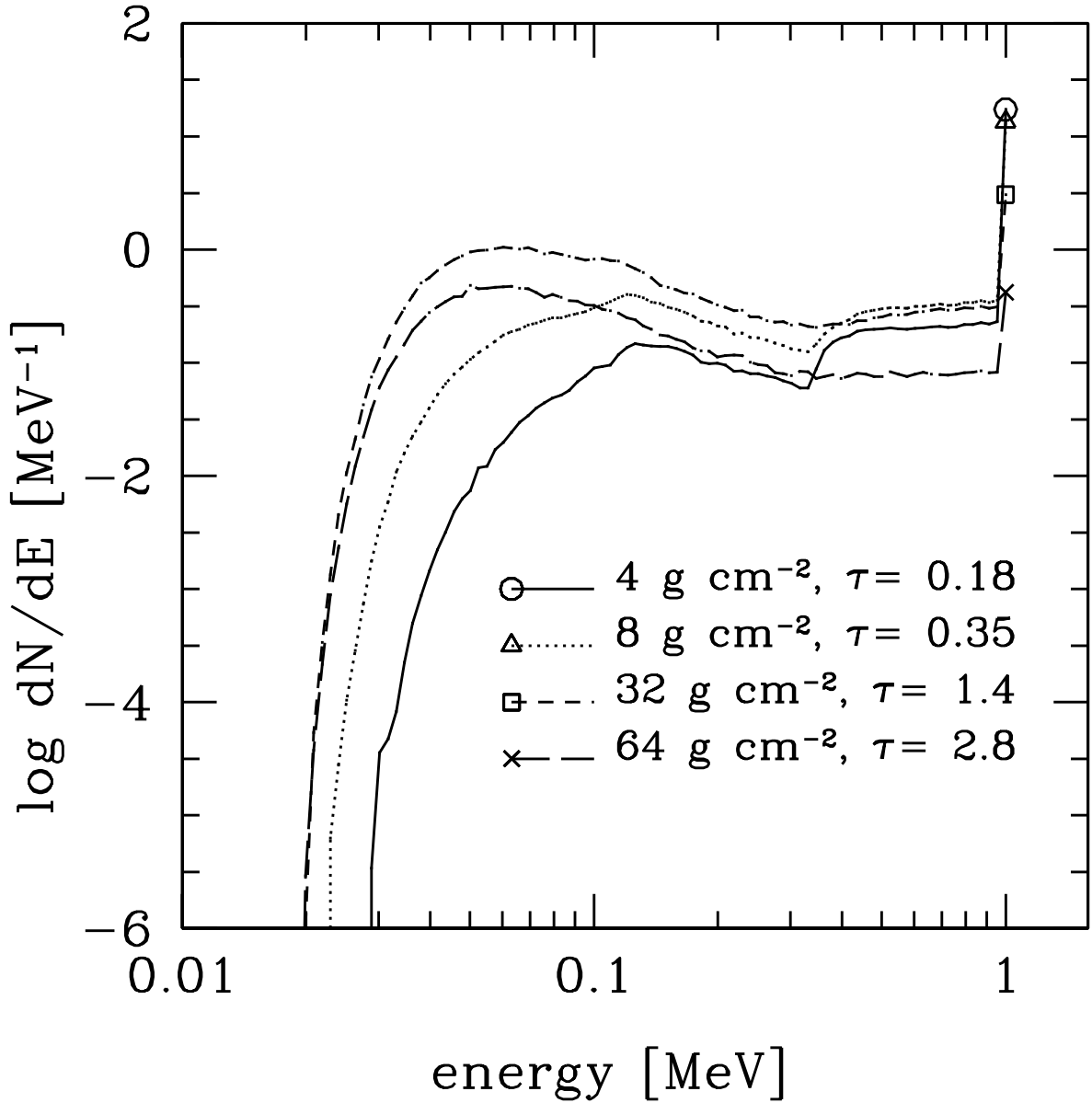} }
\caption{\label{fig:monoPhotonSpectra}Spectra of the ionizing
radiation received at the ground for four thin atmospheres and a 1
MeV monoenergetic incident spectrum.  The Compton backscattering
peaks for the first and second scatterings starting at 1 MeV can
be seen for the two thinnest atmospheres, as well as the ``piling
up'' at 50--100 keV due to successively smaller energy shifts.}
\end{figure}

Even in thin atmospheres, the surficial spectrum will be altered
from the incident spectrum.  The relevant physics can be
summarized as follows.  A photon will typically start with a very
high energy (1 MeV, say) and lose a significant fraction of its
energy to Compton recoil electrons at each interaction. This
fraction depends on energy and decreases with decreasing energy,
leading to a ``pile-up'' of photons at energies of $\lesssim 100$
keV. In the low-energy limit ($E \ll m_ec^2$), the average energy
shifts are well-approximated by $\avg{\Delta E} \simeq E^2$
(obtained by averaging the Compton energy losses weighted by the
angular Klein-Nishina cross section). The photons at successively
lower energies experience a photoabsorption cross section that
increases rapidly ($\sigma_{\rm pa}\propto E^{-3}$) and are thus
removed from the Compton downscattering peak.  This is
demonstrated in Fig.~\ref{fig:monoPhotonSpectra}, which shows the
energy spectra at the surface for four different column densities
and an incident energy of 1 MeV. The downscatter ledge, where the
photons are ``piling up,'' can be seen ($\sim 50$--100 keV) along
with a continuum of energies between this peak and the maximum
incident energy. This continuum is filled by photons that lost a
smaller than average energy at one or more Compton scatterings.
Also seen for the thinnest atmospheres are the peaks at the
average energies corresponding to both one and two Compton
scatterings from an initial energy of 1 MeV.

\begin{figure}
\centerline{ \plotone{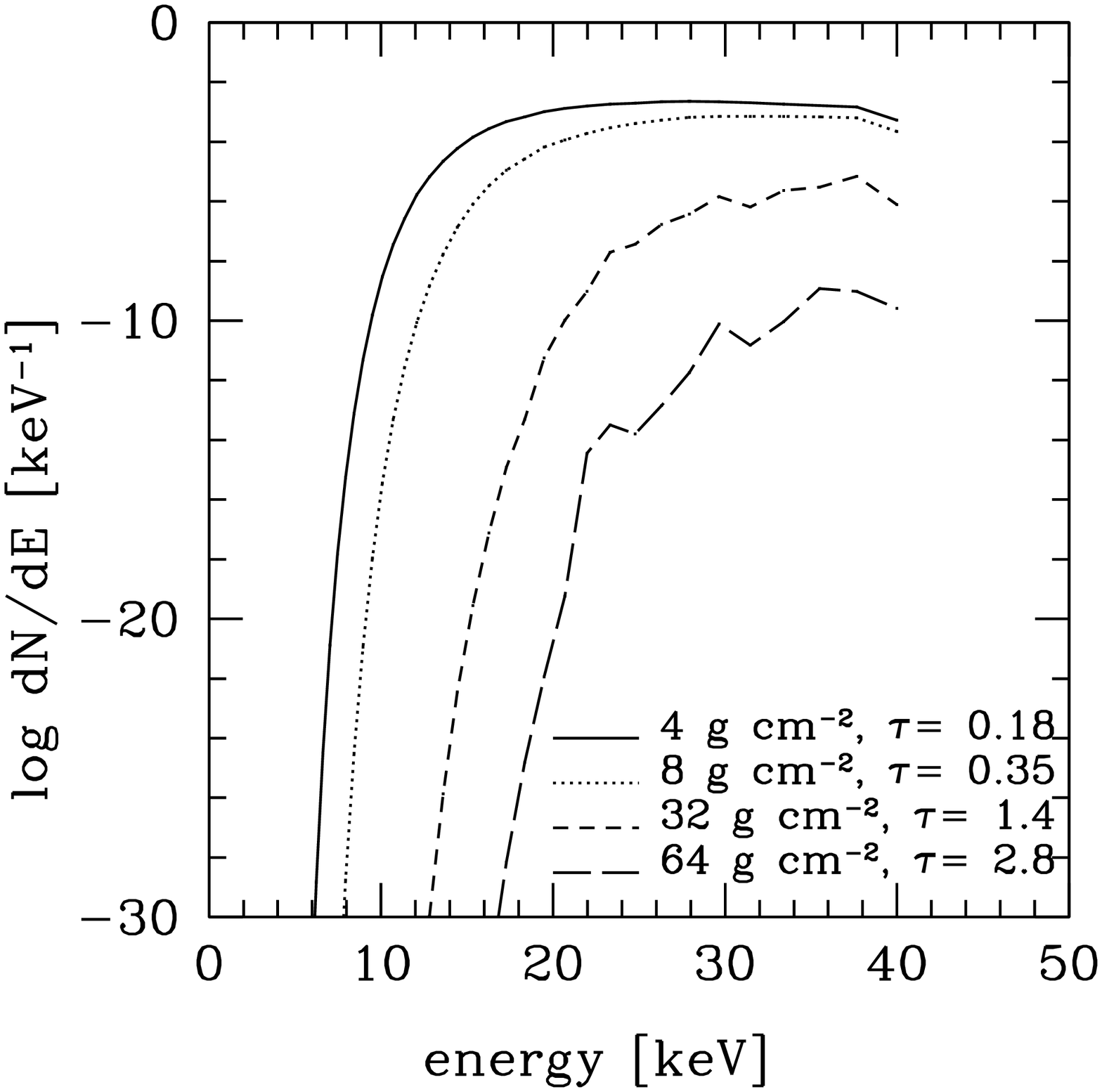} }
\caption{\label{fig:expPhotonSpectra}Spectra of the radiation
received at the ground for four thin atmospheres and an
exponential incident spectrum with an average energy of 10 keV.}
\end{figure}

\begin{figure} \centerline{ \plotone{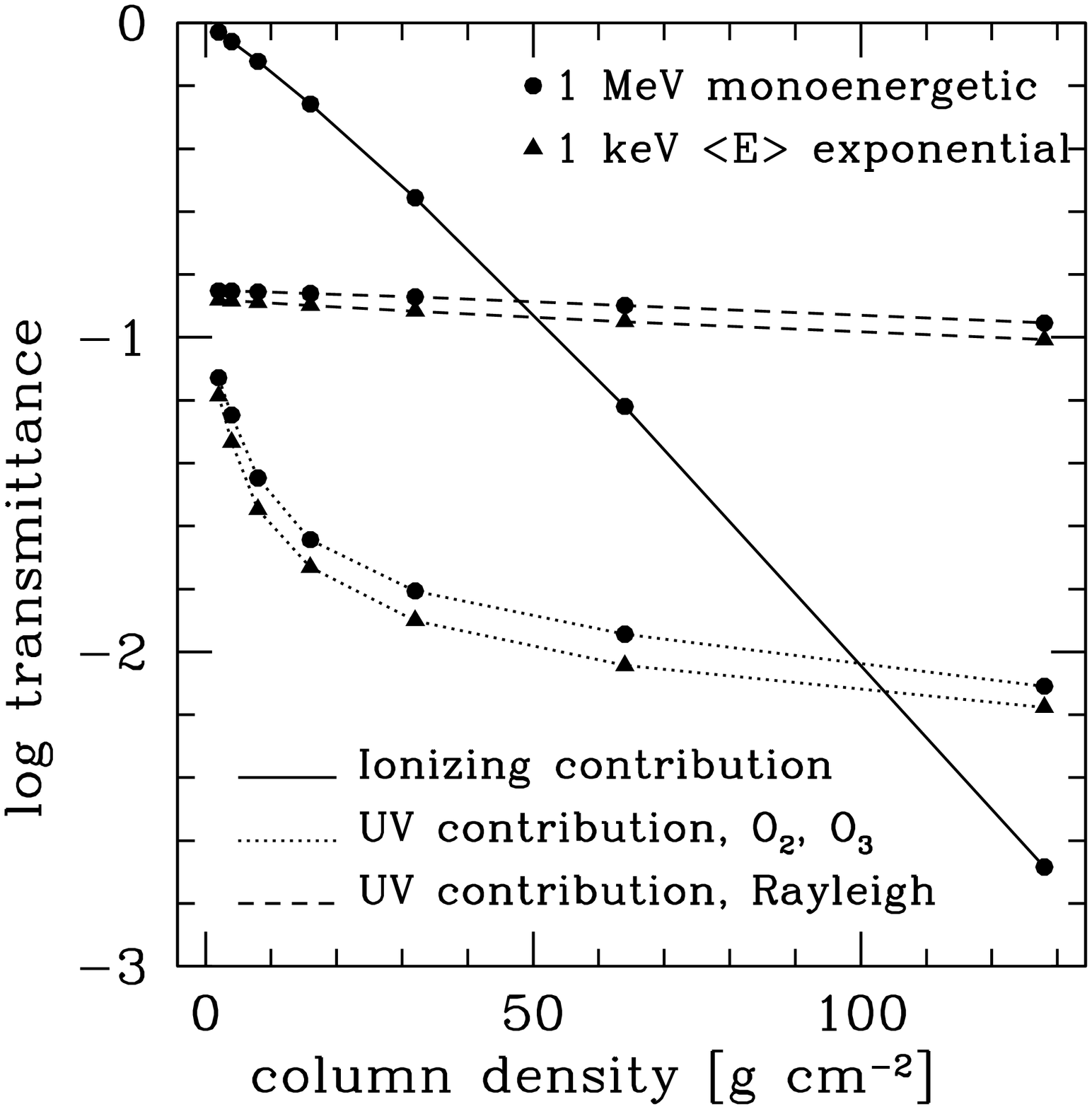}
} \caption{\label{fig:thinBioFlux}Fraction of the incident energy
reaching the ground as ionizing radiation and as biologically
effective UV in thin atmospheres for two incident spectra using
two simple models of UV redistribution.  The ionizing radiation
dominates for column densities $\lesssim$ 50 g cm$^{-2}$ for hard
incident radiation.  For the softer, $\avg{E} = 1$ keV, case,
photoabsorption prevents any substantial direct surficial flux; we
omit the solid curve corresponding to the X-ray incident spectrum
because the transmittance is far below the scale shown here. The
UV reemission contribution is shown for two cases: (1) O$_2$ and
O$_3$ molecular absorption only and (2) Rayleigh scattering only.
In both UV cases, only the biologically effective flux (200--320
nm) is counted. }
\end{figure}

For stellar flare spectra, energies are in the keV range, and the
dominant cross section is photoabsorption, and no downscattering
occurs.  Figure \ref{fig:expPhotonSpectra} shows the effect of
atmospheric attenuation on an incident stellar flare model spectrum,
represented by a decaying exponential with average energy of 10
keV.  Unlike the $\gamma$-ray case, the X-ray flare spectrum actually
shifts to higher mean energies than the incident spectrum because
the flux is attenuated primarily according to the photoabsorption
cross section, which is proportional to $E^{-3}$. The estimated
fraction of ionizing radiation received at the ground for exoplanets
with thin atmospheres is shown in Fig.~\ref{fig:thinBioFlux}.
Interestingly, since the surficial radiation spectrum has such high
characteristic energies for the thinnest atmospheres, the transmittance
is nearly independent of atmospheric composition (i.e., the primary
photon reprocessing occurs through Compton scattering, which is
independent of composition).  This makes our results quite general
if indeed exoplanets with such thin atmospheres are habitable.

Even in the optically thin limit, a finite amount of energy
redistribution to the UV occurs.  As stated above, the
$\gamma$-rays dominate the surface spectrum for atmospheres with
column densities between 30 and 50 g cm$^{-2}$; above this range,
the UV reemission dominates on planets without an atmospheric
screen. Figure \ref{fig:thinBioFlux} shows the relative
contributions of the incident radiation and the reemission to the
fraction of incident energy received at the ground in the two
extreme cases of Rayleigh scattering and absorption by an
O$_2$/O$_3$ UV shield similar to the terrestrial O$_3$
distribution with optical depth scaled as described earlier. Even
when subjected to an O$_2$/O$_3$ screen, the transmitted UV
reemission still exceeds the directly transmitted ionizing
radiation for column densities above about 100 g cm$^{-2}$, and
the transmitted fraction is about 1\% at that column density.

\subsection{Secondary ionospheric layers can be produced}

To justify our neglect of ion recombination on generic
terrestrial-like exoplanets, we must examine the most extreme
cases of irradiation. A supernova at a distance of 1 pc---which
should occur very rarely, if at all, during the lifetime of a
planetary system---would yield a maximum fluence of about 10$^8$
erg cm$^{-2}$ of ionizing radiation (hard UV and X-rays from shock
breakout and $\gamma$-ray lines); a 10$^{35}$ erg superflare of a
solar-type star would give a slightly smaller fluence for a planet
at 1 AU, while a 10$^{34}$ erg dMe flare gives a somewhat larger
fluence for a planet in the conventional habitable zone ($\sim
0.1$ AU distant for such a low mass star). These extreme events
would generate electron fractions smaller than the limit given
above in Eq.~\ref{eq:maxIonFrac}, even neglecting recombination.
The vast majority of events will easily satisfy that strong
inequality, especially if recombination timescales are not much
larger than the duration of the irradiation events. We are thus
able to obtain a reliable estimate of the maximum ionization
fractions caused by astrophysical irradiation while neglecting
recombination.

\begin{figure} \centerline{ \plotone{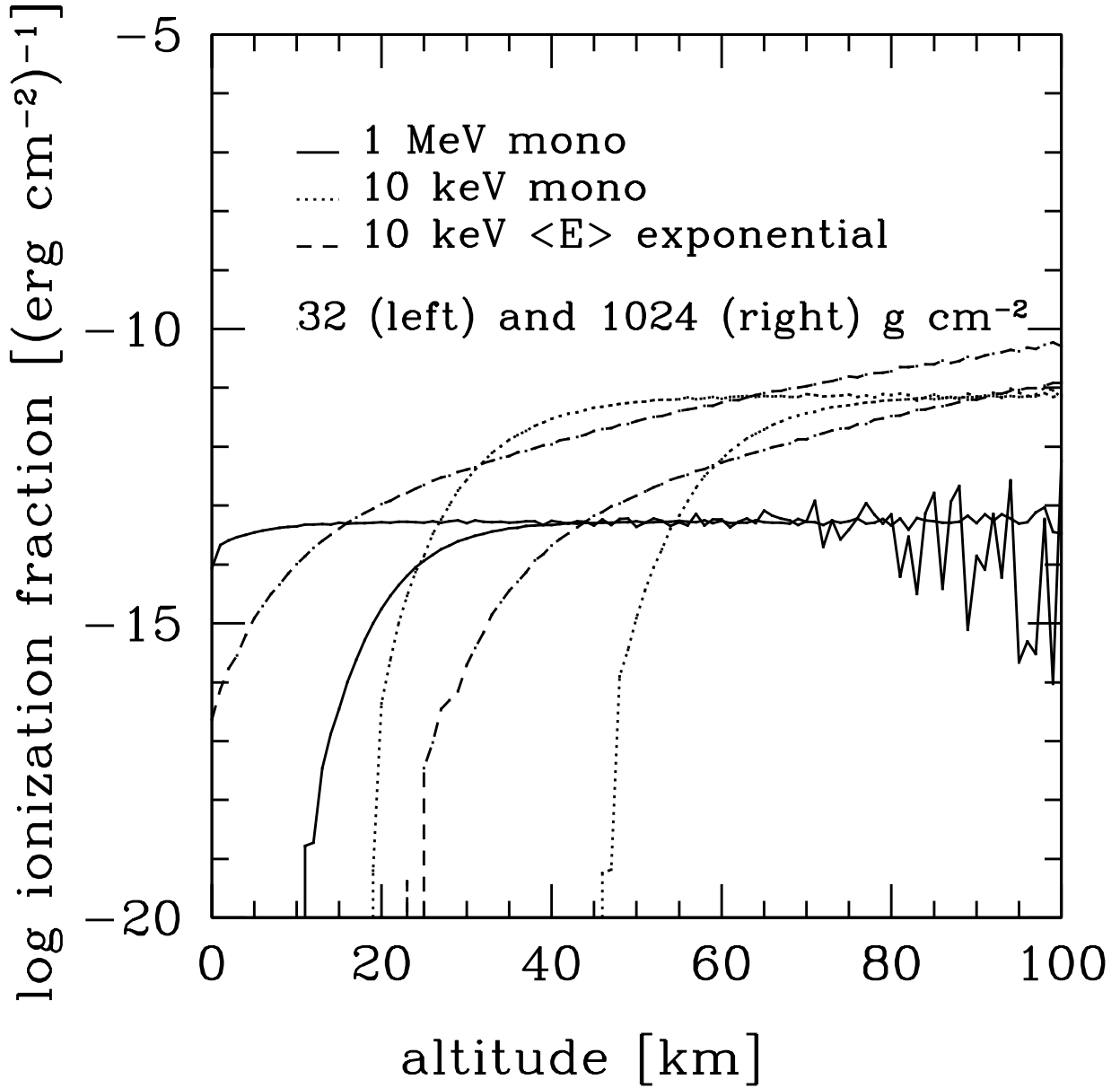}
} \caption{\label{fig:ionProfile}Typical ionization efficiency
profiles (neglecting recombination) for three different incident
spectra and two column densities. Vertical units are ionization fraction
produced per cm$^3$ per unit incident fluence at the top of the atmosphere.  For monoenergetic spectra,
nearly constant ionization levels are produced down to a
characteristic altitude, which roughly corresponds to the altitude
of maximum energy deposition. This agrees well with the Chapman
solution. The model flare spectra produce ionization levels which
rise with increasing altitude because more energy is deposited at
higher altitudes than in the monoenergetic case with an identical
average energy . The progressively larger fluctuations at
altitudes above $\sim 60$ km are due to small-number statistics,
where the optical depths are small and photon interactions are
statistically unlikely.}
\end{figure}

\begin{figure} \centerline{ \plotone{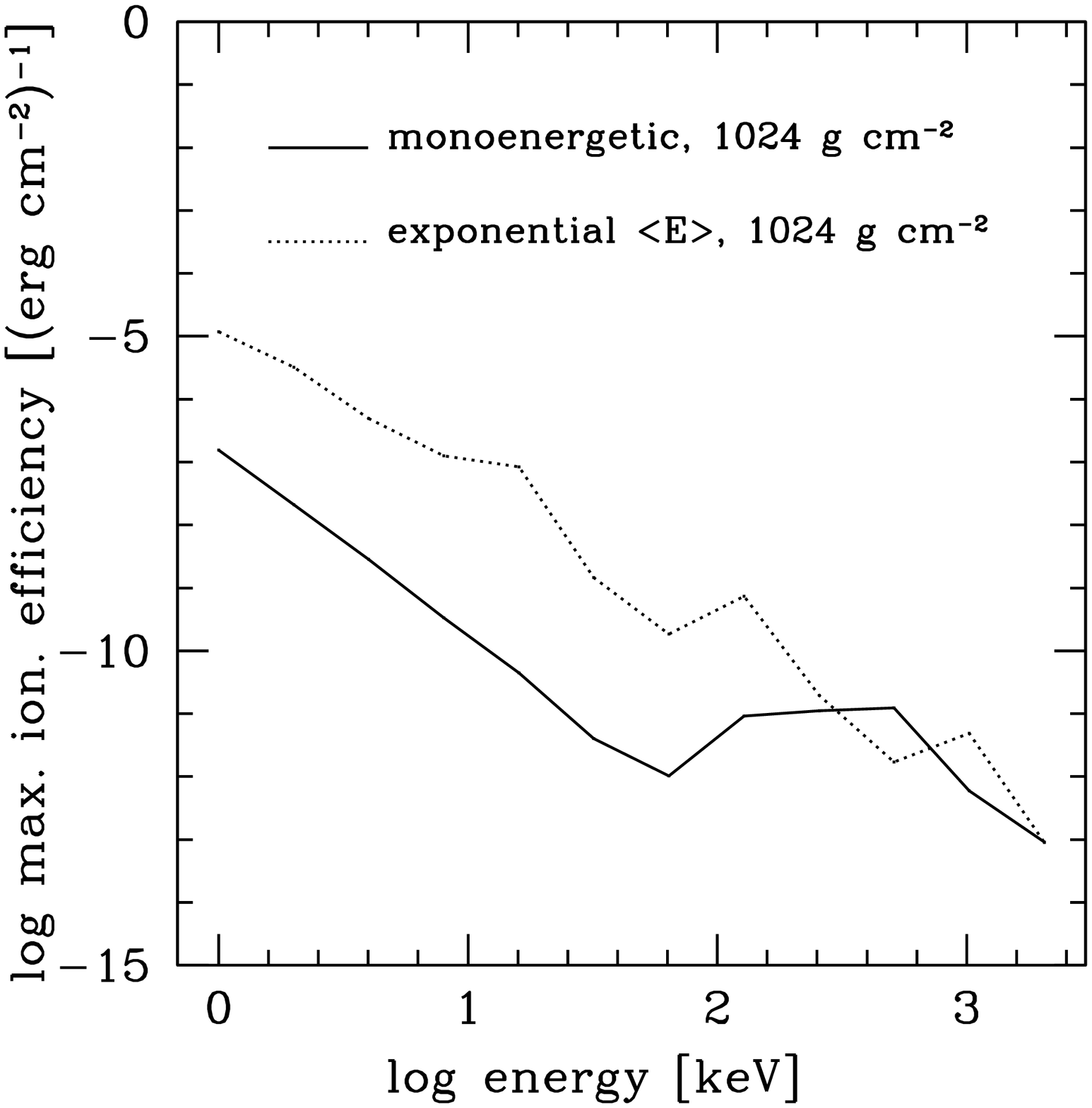}
} \caption{\label{fig:maxIonFrac}Maximum ionization efficiencies
(fractional ionization produced per unit fluence of incident
energy) for average incident energies between 1 keV and 2 MeV and
a column density of 1024 g cm$^{-2}$. The efficiency declines as
energy increases because most of the energy is being deposited at
lower altitudes where molecular number densities are higher and
hence a given amount of energy is able to ionize a smaller
fraction of the molecules.  Uncertainties in the curve arise from
small-number statistics---only a small fraction of the incident
photons are interact in the layer of maximum energy deposition,
creating fluctuations in the altitude of maximum ionization.}
\end{figure}

Even with moderate levels of irradiation, regions of the terrestrial-like
exoplanet atmosphere can be ionized to the level of the terrestrial
ionosphere.  For comparison, the ionization fractions in the
terrestrial ionospheric D and E layers are \sci{1}{-12} and \sci{1}{-7}
\citep{crisp00}, respectively.  The ionization profiles we calculate
show nearly constant ionization levels for the monoenergetic spectra
down to a characteristic altitude, below which the ionization level
drops extremely rapidly.  For our continuous spectra, ionization
levels rise with increasing altitude because the photon number per
unit energy decreases with increasing energy.  Thus most photons
in our continuous spectra have lower energies (and larger interaction
cross sections) than the spectrum average and will be deposited at
higher altitudes.  Figure \ref{fig:ionProfile} shows the results
for a few atmosphere models. Even for atmospheres as thick as
Earth's, the ionization profiles shown as fractional ionization
produced per unit incident fluence for $\gamma$-ray incident spectra
show a significant effect down to altitudes below the lowest
steady-state ionization layer on Earth (D layer, 60--95 km;
\citealt{crisp00}). For example, fluences at the top of the atmosphere
of 1 erg cm$^{-2}$ from a stellar flare with average energy of 10
keV would be yield a transient layer of comparable ionization
fraction to the D layer, but at much lower altitudes.  Based on
this result, we predict that additional ionization layers may be
produced on a transient basis and with stochastic ionization levels
in response to external radiation sources. Neglecting recombination,
the maximum ionization fractions per unit incident fluence (hereafter,
ionization efficiency) are independent of column density, depending
only on the incident spectrum. For stellar flare irradiation with
hard X-ray spectra of average energy in the range 1--10 keV, we
find maximum ionization efficiencies of 10$^{-5}$--10$^{-7}$ (erg
cm$^{-2}$)$^{-1}$; for supernovae and gamma-ray bursts, we find
maximum ionization efficiencies of 10$^{-12}$--10$^{-13}$ (erg
cm$^{-2}$)$^{-1}$. Maximum ionization efficiencies as a function
of average incident energy are shown in Fig.~\ref{fig:maxIonFrac}.

\begin{figure}
\centerline{ \plotone{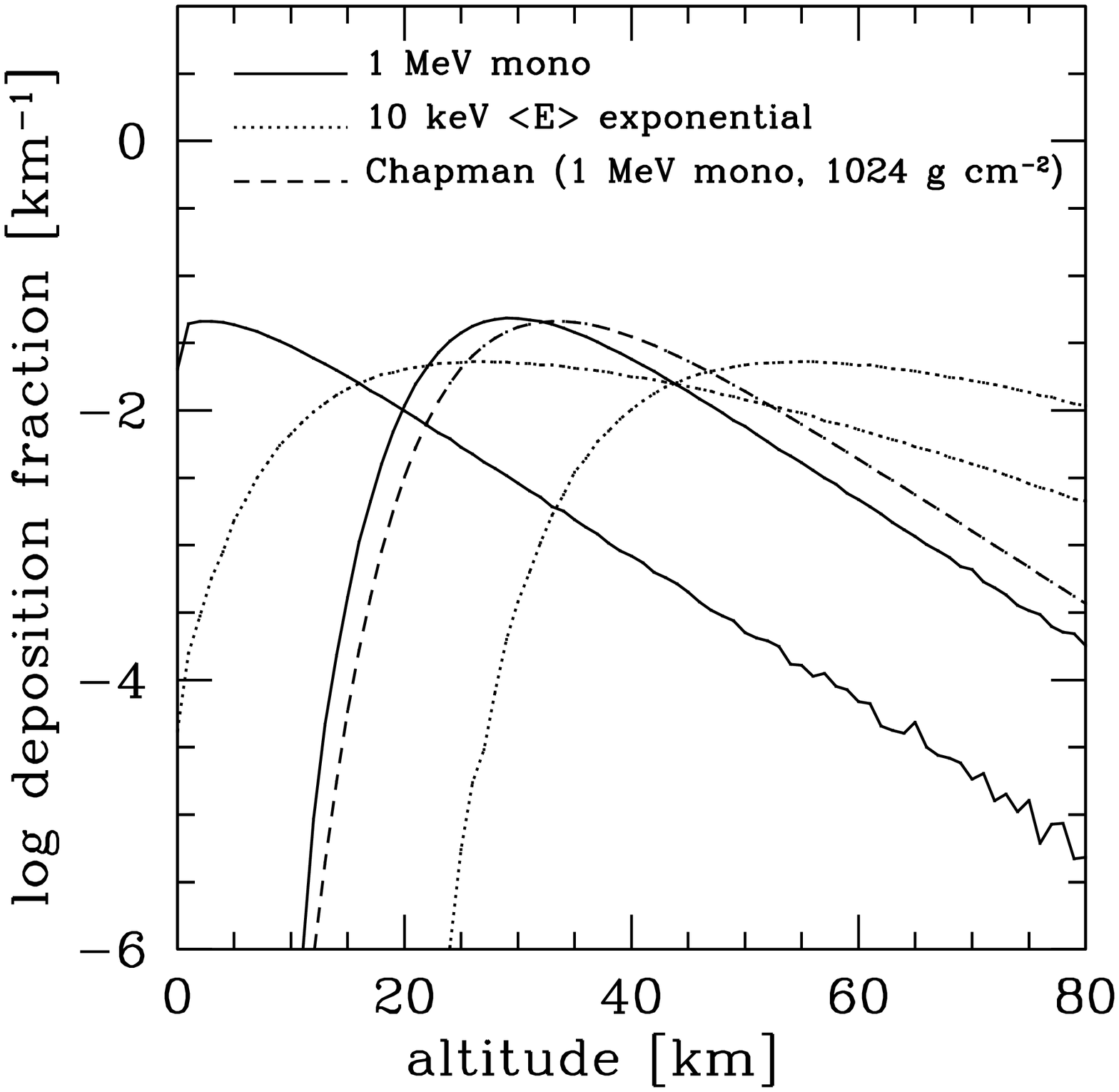} }
\caption{\label{fig:energyDepProfile}Energy deposition profiles
for three different incident spectra in atmospheres with column
densities of 32 and 1024 g cm$^{-2}$.  For each spectrum, the
lefthand curve is for a 32 g cm$^{-2}$ column density and the
righthand curve is for a 1024 g $^{-2}$ column density. A single
curve corresponding to the Chapman solution for a 1 MeV
monoenergetic incident spectrum in a 1024 g cm$^{-2}$ atmosphere
is shown for comparison. Note that our Monte Carlo results for the
monoenergetic spectra are identical in shape to the Chapman
solution (which assumes a monoenergetic incident spectrum) except
for a shift to lower altitudes which accounts for the effects of
multiple scattering. Also it can be seen that continuous spectra
give different overall shapes, with our model flare spectrum
depositing more energy higher in the atmosphere than a
monoenergetic spectrum at the same average energy.}
\end{figure}

As the ionization fraction at a particular layer depends on the
amount of energy deposited in that layer, it is instructive to
examine the energy deposition profiles, defined here as the variation
in the fraction per km of the incident energy transferred to
photoelectrons and Compton-recoil electrons as a function of altitude.
Figure \ref{fig:energyDepProfile} shows that our model stellar flare
spectrum deposits more energy at higher altitudes than the corresponding
monoenergetic incident spectrum at the same average energy.  This
has important implications for the ionization fractions created.
Because the density of atmospheric molecules falls off exponentially
with height, a higher fraction of energy deposited at higher altitudes
where molecular densities are lower will result in higher ionization
fractions.  This is why the stellar flares may create higher peak
ionization fractions than supernovae, and why the altitudes of peak
ionization will be higher.

\subsection{Substantial diffuse UV is produced in thick
atmospheres} \label{sec:uvthick}

We can estimate the intensity of aurora-like emission produced by
superflares and cosmic explosions here and on other planets by
comparing with terrestrial data.  Auroral intensities are often
classified into four ``International Brightness Coefficient''
(IBC) Classes I-IV, from weakest to strongest.  According to data
presented by \citet{whalen85} for IBC Class III auroral
intensities, the OI 557.7 nm emission is about 1\% of the total
zenith auroral brightness. For the most intense auroral events
with IBC Class IV the brightness in the OI line is 10$^{12}$
photons cm$^{-2}$ s$^{-1}$. Using the same scaling from OI to
total brightness as for the Class III event, the total brightness
must be of order $10^{14}$ photons cm$^{-2}$ s$^{-1}$.  Using 5 eV
as a median energy photon for the auroral emission, this gives a
rough energy flux of \sci{8}{2} erg cm$^{-2}$ s$^{-1}$.  For the
Class III data, the efficiency of conversion of primary and
secondary electron energy into radiation at all wavelengths is
given to be 21--35\%, so the corresponding photon flux for the
Class IV event is about \sci{3}{3} erg cm$^2$ s$^{-1}$.  We find
larger efficiencies for the ratio of incident photon energy to
electron energy for the very different physical process producing
the electrons here (Compton scattering and photoabsorption, versus
collisional ionization for standard aurorae), and similar
efficiencies can be inferred from calculations of X-ray
redistribution in accretion disks around compact stellar objects
\citep{ross79,kallmanMccray82,rossFabian93}.

In comparison to these terrestrial events, we have estimated that
gamma-ray burst events \citep{scaloWheeler02} and supernova
explosions \citep{scalo03} would expose an exoplanet to incident
ionizing fluences greater than 10$^6$ erg cm$^{-2}$ hundreds of
times per Gyr, which translates to fluxes of about 10$^7$ erg
cm$^{-2}$ s$^{-1}$ and 10$^2$ erg cm$^{-2}$ s$^{-1}$,
respectively.  On a planet orbiting a low-mass dMe strong flare
star in the habitable zone (semimajor axis $\sim 0.1$ AU), a flare
with an EUV energy greater than 10$^{32}$ erg can occur 10--100
times per day (see \citealt{audard00}, Fig.~4), with a
corresponding flux for a 10 minute flare of 6000 erg cm$^{-2}$
s$^{-1}$. Given the energy-frequency power law relations estimated
for both solar (e.g., \citealt{crosby93} and
\citealt{aschwanden00}) and dMe flares in various UV and X-ray
bands \citep{gershbergShakhovskaya83,audard00,guedel03}, with
differential frequency distributions of $-1.5$ to $-2.2$, incident
fluxes of at least 10$^5$ erg cm$^{-2}$ s$^{-1}$ should occur with
a frequency of order once per day. Clearly the intensities of
auroral lines generated by these events will far exceed the
strongest terrestrial Class IV auroral displays.

\begin{figure} \centerline{ \plotone{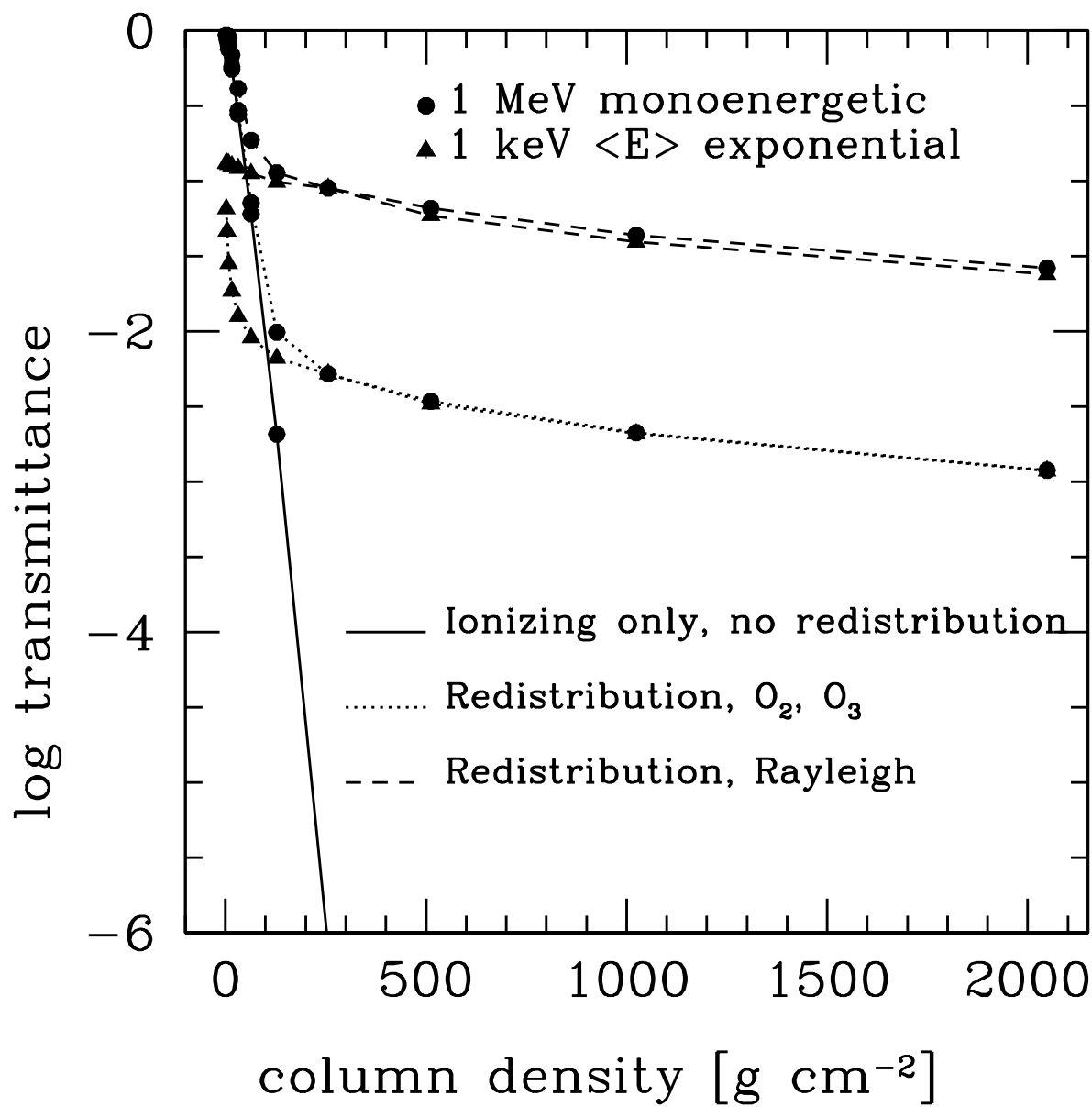}
} \caption{\label{fig:bioFlux} Fraction of the incident energy
reaching the ground is shown both without and with the additional
contribution of the redistributed UV for two simple models of UV
redistribution for an atmosphere with a column density of 1024 g
cm$^{-2}$. The no redistribution case for a 1 keV average energy
exponential spectrum is not shown because the transmitted fraction
is practically zero.  } \end{figure}

Figure \ref{fig:bioFlux} shows the fraction of the incident fluence
reaching the surface in the biologically significant range 200--320
nm for column densities up to 2048 g cm$^{-2}$. The original incident
radiation is strongly attenuated, but the redistribution of energy
toward UV emission maintains the surface fluences at significant
levels. Two cases are shown: pure Rayleigh scattering and pure
O$_2$/O$_3$ absorption.  The pure Rayleigh scattering case represents
an atmosphere with no significant molecular or aerosol UV absorbers
in the biologically effective region---perhaps similar to the
Archaean Earth. The transmission in this case was calculated using
the modified two-stream \citet{schuster05} scattering solution
described in Appendix \ref{sec:schusterModification}. The O$_2$/O$_3$
case represents an ozone and oxygen abundance similar to the
present-day Earth (identical column density profiles).  As can be
seen in Fig.~\ref{fig:bioFlux}, the effect of redistribution to the
UV is quite dramatic, even when subjected to molecular absorption
by O$_2$ and O$_3$. The UV reemission quite effectively raises the
surficial fluences back to significant levels, even though the
incident ionizing radiation has been attenuated to ridiculously
small amounts in the thick atmospheres. For example, the fraction
of incident X-rays and $\gamma$-rays reaching the surface on Earth
(1024 g cm$^{-2}$) is \sci{6}{-29} for the 1 MeV monoenergetic case,
while including the UV redistribution to the biologically relevant
200--320 nm region, even in the presence of an ozone screen, raises
this number to at least \sci{2}{-3}.

\begin{figure} \centerline{ \plotone{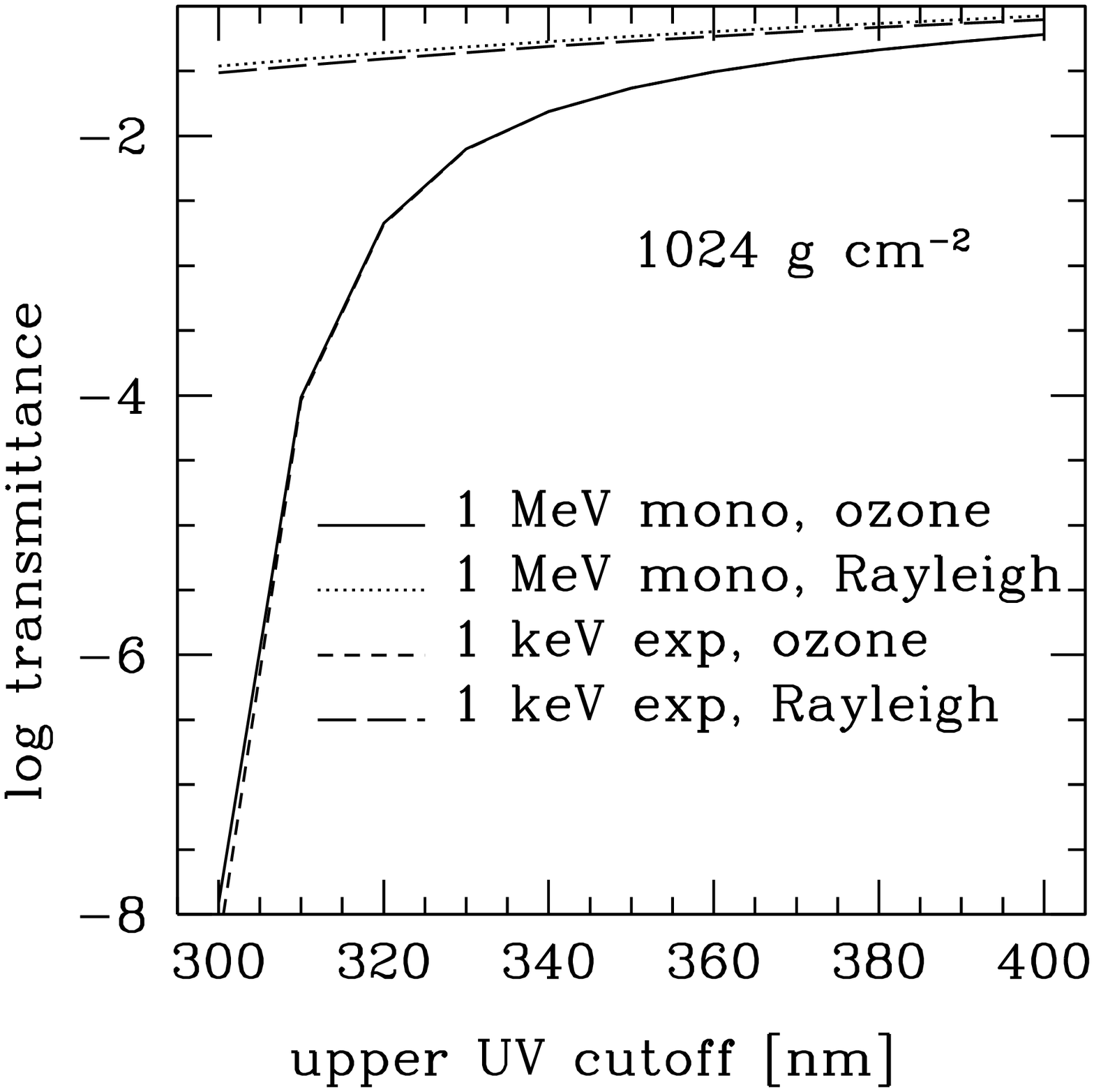} }
\caption{\label{fig:uvCutoff}Fraction of biologically significant
redistributed UV energy that is reaching the ground on Earth as a
function of the upper cutoff for the biologically significant flux.
The lower limit is taken to be 200 nm in all cases.  The result is
insensitive to the cutoff for the Rayleigh scattering case, but
very sensitive in the O$_3$ case for cutoffs in the range 300--340
nm.  The sharp falloff in the O$_3$ case for wavelengths shorter
than 340 nm is due to the rapidly increasing O$_3$ cross section
at smaller wavelengths (peaking around 260 nm).  Significant
biological effect occurs up to 350 nm in some organisms under UV
irradiation.  Our work assumes 320 nm for the cutoff.  } \end{figure}

The results depicted in Fig.~\ref{fig:bioFlux} depend sensitively
on the adopted upper limit of 320 nm for ``biologically
effective'' UV radiation, but we feel that a value of 320 nm is
quite reasonable. Figure \ref{fig:uvCutoff} shows the strong
dependence of the transmission on the adopted upper wavelength
limit for ``biological significance.'' For the case of O$_3$
absorption, the dependence is quite severe, so the transmission
will depend on which specific biological process is of interest.
It is well known that UV-B radiation around 320 nm has major
effects on contemporary organisms and ecosystems, and even
wavelengths as large as 350 nm can have a variety of biological
effects (e.g., \citealt{jagger85}). For example the action
spectrum for induction of squamous cell carcinoma in mice has a
strong peak at 300 nm and is smaller by only an order of magnitude
at 320 nm (\citealt{nilsson96}, p.~88).  Additionally, UV-B
(280--315 nm by convention) can penetrate ocean surfaces to much
larger depths than UV-C (100--280 nm).  Although many DNA action
spectra peak at 260 nm and have declined by a factor of 10--100 by
300--320 nm, the action spectrum for particular \emph{mutations}
do not show this universal behavior.  As one of many well-known
examples, the measured action spectrum for the UV-induced mutation
to resistance to novobiocin in \emph{Haemophilus influenzae} has a
sharp peak around 330 nm and drops by two orders of magnitude
below 280 nm and above 360 nm \citep{cabrerajuarezSetlow76}. A
recent survey of the numerous effects of the longer-wavelength
UV-B radiation on terrestrial organisms and ecosystems is given by
\citet{paulGwynnjones03}.

\subsection{Additional considerations}

\subsubsection{The Chapman solution is inaccurate for thin
atmospheres and high energies}\label{sec:chapman}

The problem of energy deposition in an exponential atmosphere for
the case of pure absorption was solved by \citet{chapman31}. While
our situation is more complicated, the photons are nevertheless
depositing energy in the atmosphere, and we can compare our results
to Chapman's solution, which is commonly used to estimate the effects
of ionizing radiation (e.g., \citealt{gehrels03}).  Assuming an
exponential attenuation in an exponential atmosphere, Chapman showed
that the energy deposition rate, $q$, as a function of altitude is
(see \citealt{chamberlain78}) \beq \label{eqn:chapmanProf} q(y) =
q_\mathrm{max} \exp\left(1 - y - e^{-y}\right), \eeq where
$y=(z-z_\mathrm{max})/h$ is the dimensionless altitude, $z_\mathrm{max}$
is the altitude of maximum deposition, and $h$ is the atmospheric
scale height.  This solution is based upon an exponentially increasing
optical depth, so that the attenuation of the radiation (and hence
the energy deposition) follows the profile of an exponential raised
to an exponential.  Additionally, the altitude of maximum
deposition in the Chapman solution scales logarithmically with the
optical depth: \beq \label{eq:chapMaxHeight} z_\mathrm{max}=h\log\tau.
\eeq

Our Monte Carlo results, which take into account multiple scatterings,
yield an energy deposition curve which is of the same shape as that
which the Chapman solution (which assumes a monoenergetic incident
spectrum) predicts, but with the entire curve shifted to lower
altitudes due to the effects of multiple scattering. This is evident
in Fig.~\ref{fig:energyDepProfile}.  Also it can be seen that
continuous spectra give different overall shapes, with our model
flare spectrum depositing more energy higher in the atmosphere than
a monoenergetic spectrum at the same average energy. In principle,
the shape of the energy deposition curve for the continuous spectrum
could be retrieved via the Chapman solution by summing appropriately
weighted Chapman curves at each energy in the range of energies in
the continuous spectrum.

In short, the general shape of the Chapman profile is accurate at
one particular energy (and hence for our monoenergetic supernovae
spectra), but it underestimates how far into the atmosphere the
radiation will penetrate because of the neglect of multiple scattering.
This effect is minor for the thickest atmospheres (column densities
$\gtrsim 300$ g cm$^{-2}$) but becomes significant for thinner atmospheres
(column densities $\lesssim 300$ g cm$^{-2}$).

\begin{figure} \centerline{ \plotone{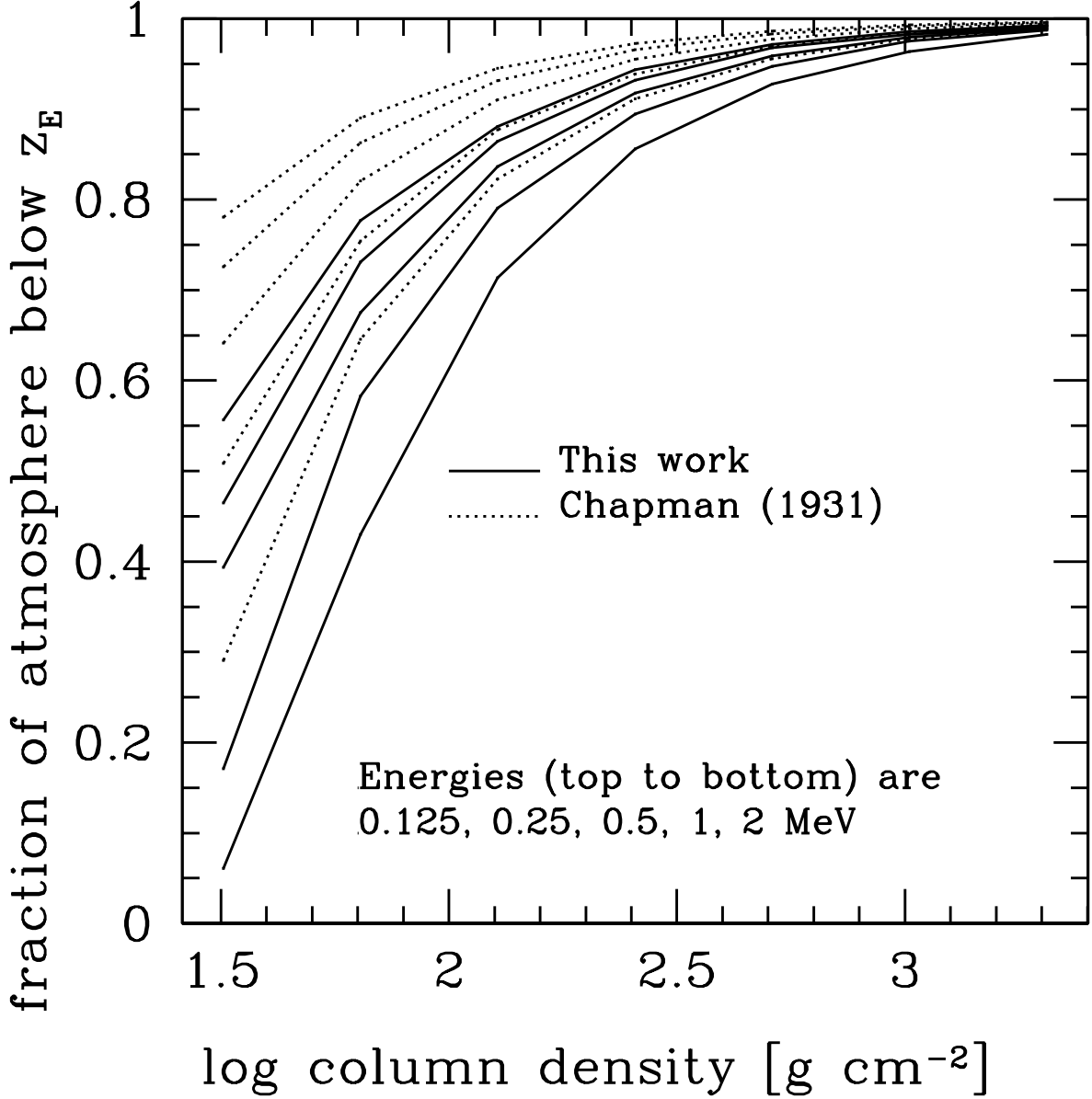}
} \caption{\label{fig:heightEDep}Comparison of the calculated
location of peak energy deposition of monoenergetic radiation with
that predicted by the Chapman solution. Results are presented in
terms of the fraction of the atmosphere by mass penetrated by the
energy deposited at the height of maximum energy deposition. Since
the Chapman mechanism neglects subsequent scatterings, the altitudes
of maximum energy depositions are higher than in our calculations. The
effect is quite significant for the thin atmospheres with column
densities $\lesssim 100$ g cm$^{-2}$.} \end{figure}

To gauge the effect of multiple scatterings in our Monte Carlo
model, we calculated (Fig.~\ref{fig:heightEDep}) the fraction of
the atmosphere above the height of maximum energy deposition,
which is a measure of how far the radiation has penetrated the
atmosphere. The effects of lower deposition altitudes are more
pronounced for thinner atmospheres and higher energies.  More
scatterings occur before the photons are photoabsorbed for higher
incident energies, and each scattering has a longer mean free path
in thinner atmospheres. In terms of the fraction of the
atmospheric mass penetrated by the radiation, the full radiative
transfer yields 10--50\% greater penetration, depending on the
thickness of the atmosphere and the energy of the incident
radiation.  At low energies the photoabsorption dominates, so the
results approach the Chapman solution.  The Chapman monoenergetic
solution could be used to build up solutions for continous
spectra, so it is not invalid for incident spectra such as flares or
gamma-ray bursts.  The key quantity is the average energy
of the spectrum and the thickness of the atmosphere. The most
pathological situation for the Chapman solution is a high-energy
radiation source (e.g., supernovae and gamma-ray bursts) incident on a thin
atmosphere ($\lesssim$100 g cm$^{-2}$).

\subsubsection{Surface energy fluences are not sensitive to the angle of
incidence} \label{sec:angInc}

\begin{figure} \centerline{ \plotone{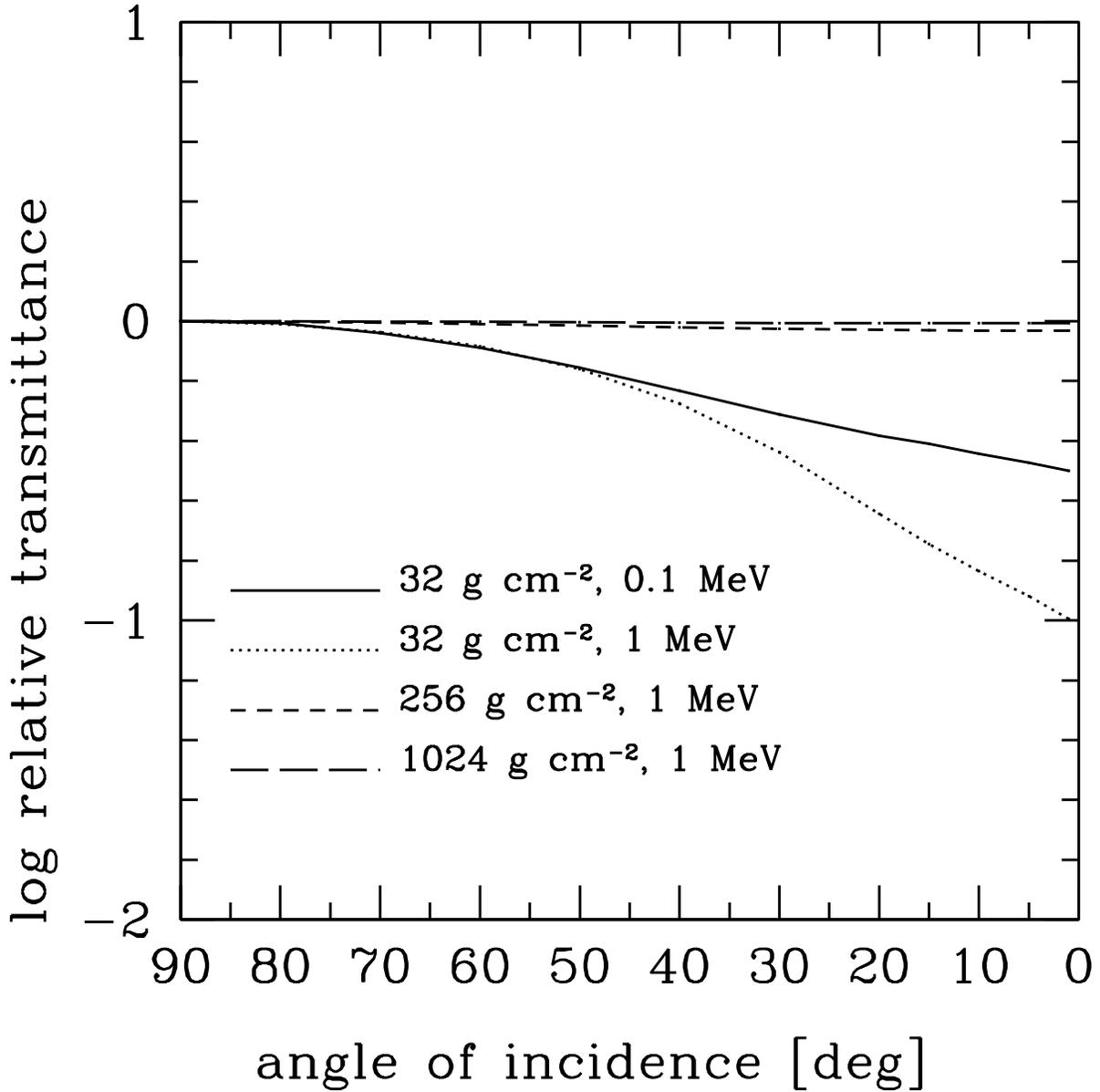} }
\caption{\label{fig:angleOfIncidence}The effect of varying the angle
of incidence. Plotted is the transmittance of the atmosphere as a
function of angle of incidence relative to normal incidence.  The
curves include UV redistribution subjected to O$_2$/O$_3$ absorption
in the transmittance (which raises the transmitted fractions above that
of only the direct ionizing radiation). In the thick atmospheres,
we find that the angle of incidence has only a very small effect
on the reemitted UV that reaches the surface when redistribution
is included; for thin atmospheres (and hence small optical depths),
the effect is negligible for both the incident radiation and the
redistributed radiation.} \end{figure}

We find that decreasing the angle of incidence can measurably
decrease the surface fluence of the \emph{original} incident
radiation, with the effect becoming quite significant for the
highest energies ($\gtrsim$ 1 MeV) and thickest atmospheres
($\gtrsim$ 500 g cm$^{-2}$). Though in our model we assume normal
incidence for every photon, the surface of a real exoplanet
subjected to a source at astronomical distances will observe a
point source with varying zenith distances, depending on the
viewing geometry.  This effect means that we have
calculated only an upper limit to the direct transmittance of the
atmospheres to the incident \emph{ionizing} radiation.  We note that
this ionizing radiation is already insignificant in a biological
sense for column densities greater than about 100 g cm$^{-2}$. As
found earlier, the primary contribution to the surface flux in
thick atmospheres is the redistributed UV. Since the UV is
primarily attenuated by molecular absorbers, atmospheres thick
enough to deposit most of the incident energy above the absorbers
will be indifferent to the angle of incidence of that radiation.
Figure \ref{fig:angleOfIncidence} illustrates the effect. In the
thick atmospheres, we find that the angle of incidence has only a
very small effect on the reemitted UV that reaches the surface in
the biologically effective region of the spectrum; for thin
atmospheres, the effect is negligible for both the incident
radiation and the reemitted radiation because the optical depths
are by definition small. We are therefore justified in neglecting
the effects of the angle of incidence.

\section{Relevance of results to astronomical sources}\label{sec:srcs}

Our work is based on the fact that most planetary systems must be
occasionally irradiated by bursts of X-ray and $\gamma$-ray photons
from various astronomical events, a facet of planetary radiation
environments that has been overlooked in the past.  In order to
place the above calculations in their proper context, we briefly
summarize the major sources of such ionizing radiation, concentrating
on the estimated frequencies and durations of the events.

\subsection{Stellar flares}\label{sec:srcs_flares}

Surely the most frequent sources of stochastic irradiation by
ionizing photons are flares from the parent star.  For older,
solar-like stars, the Sun provides the best-studied example. Solar
flares are associated with ionizing radiation from keV X-rays to
GeV $\gamma$-rays. The time variation of flare output depends on
wavelength region and varies from flare to flare, e.g., Fig.~10.11
in \citet{foukal90}, Fig.~6.7 in \citet{lang00}, and Fig.~9.2 in
\citet{golubPasachoff97}, with the gradually declining flare phase
lasting up to an hour or more.  The radiative energy release in a
single flare varies by orders of magnitude, with the strongest solar
flares ever observed emitting a few times $10^{32}$ erg (e.g., the
4 Nov 2003  flare).  This corresponds to a flux above the Earth's
atmosphere of only $60\ E_{32}$ erg cm$^{-2}$ s$^{-1}$, where
$E_{32}$ is the flare energy in units of 10$^{32}$ erg and we adopted
an average duration of 10 min.  This is consistent with the maximum
observed soft X-ray peak fluxes (Fig.~1 in \citealt{zirin00}).

Given the small historical interval over which such observations
are available, even in the visual part of the spectrum, and the
rapidly decreasing frequency of higher-energy events, it is reasonable
to assume that still higher-energy events do occur, even if they
have not been detected. The frequency distribution of flare energies
from EUV to hard X-rays, derived from several space missions, can
be described by a fairly robust power law
(\citealt{crosby98,aschwanden00,lin01,guedel03}, and references
therein), with log-log slope about $-1.6$. Extrapolating these data
to higher energies, we find that the frequency of flares of such
large energy that 1\% of the X-ray energy (using our result for an
atmosphere of column density similar to the Earth's) exceeds the
solar UV flux at 1 AU of about \sci{1}{4} erg cm$^{-2}$ s$^{-1}$
at the Earth's surface in the biologically active 200--320 nm region
is about one per century.  This frequency should be decreased if
the steeper soft X-ray (0.1--0.8 nm) energy-frequency distribution
recently found by \citet{veronig02} is correct.

That such flares do occur, even in old, weak-flaring stars like the
Sun, is supported by \citet{schaefer00}, who have identified nine
cases of superflares with energy outputs of $10^{33}$--$10^{38}$
erg on otherwise normal F8--G8 main sequence stars.  These flares
cannot be attributed to binaries, rapid rotation, or youth, and
therefore may be common in solar-type stars. \citet{schaefer00}
additionally estimate a very uncertain recurrence time of $10^2$--$10^3$
yr.

Intense flares are much more frequent in younger solar-mass stars,
as evidenced by both coronal X-ray emission of solar analogues of
different ages \citep{guinanRibas02} and estimates of stellar wind
momentum fluxes from solar-mass stars of different ages \citep{wood02}.
It is also known that intense, although less frequent, flares occur
in stars even more massive than the Sun, such as the F star EUV
flares observed by \citet{mullanMathioudakis00}.

Using the available data on solar-mass stars of various ages, we
find that the frequency of flares energetic enough to yield, after
redistribution, UV fluxes in excess of the stellar flux should be
on the order of once per 1--1000 yr depending on stellar age for
habitable zone planets with atmospheres as thick as that of the
Earth orbiting solar-mass and higher-mass stars. The frequency will
be larger for planets with smaller atmospheric column densities;
the dependence of the fraction of flux redistributed from X-rays
to UV as a function of planetary atmosphere column density is shown
in Fig.~\ref{fig:bioFlux}. Although even small changes in the UV
irradiance can have sizeable effects on the Earth's atmosphere (see
\citealt{larkin00}), the brevity and large duty cycle of very
energetic flares make their importance uncertain for solar-mass
stars.

The situation is quite different for lower-mass, red main sequence
stars of spectral type M. These stars are the most numerous stars
in the Galaxy (see \citealt{chabrier03} and references therein),
and calculations indicate that atmospheric circulation is sufficient
for atmospheric retention and liquid water oceans in spite of
synchronous rotation \citep{joshi97,joshi03}.  The potential
importance of these stars for exobiology was first clearly recognized
and discussed in detail by \citet{heath99}.

Very low-mass stars spend a significant fraction of their long lives
in a state dominated by strong and frequent flare activity (e.g.,
\citealt{shakhovskaya95}).  Such stars, called ``flare stars,''
``emission line stars,'' or ``UV Ceti stars'' (after the prototype),
are designated as spectral type dMe (see \citealt{gershberg99} for
an extensive database and bibliography).

These low mass stars are sources of frequent intense flares with
energies as large as $10^{34}$--$10^{35}$ erg in ionizing radiation
\citep{cully93,hawleyPettersen91} occurring roughly once per 100
hours of monitoring for some stars, with larger energies occurring
at smaller rates. Figure 4 of \citet{audard00} shows that the rate
of EUV flares with energies exceeding 10$^{32}$ erg ranges from
$\sim$ 0.1--100 per day, depending on the star's coronal X-ray
luminosity (which is correlated with age). There are several examples
of dMe stars with intense flares of energies exceeding 10$^{34}$
erg in the blue and UV, as summarized by \citet{liebert99}; see
also \citet{hawleyPettersen91} and \citet{pagano97}. Although these
cases were not observed in the X-ray region, examples exist of
comparable X-ray flares in other dMe stars (e.g., EV Lac,
\citealt{favata00}). The larger and more frequent energy releases
in very low mass star flares are accentuated by the proximity of
conventional habitable zones: $\sim$ 0.05--0.15 AU for stars in the
range of masses 0.1--0.4 $M_{\odot}$ (see \citealt{kasting93}).
Since the habitable zone distance is partly determined by bolometric
flux, habitable planets around these stars will be subjected to
flare rates and fluxes many orders of magnitude larger than the
Earth.

As a specific example, \citet{cully93} describe soft X-ray flares
of energy above 10$^{34}$ erg lasting over 2 hours for the dMe star
AU Mic.  This would give a flux above a habitable zone planet
atmosphere of about 10$^5$ erg cm$^{-2}$ s$^{-1}$.  Using the results
of \S\ref{sec:results} for the UV redistribution, and the relative
UV fluxes expected in dM stars, we find that the redistributed flare
energy would swamp the stellar photospheric UV by an order of
magnitude for a habitable planet atmosphere as thick as Earth's.
Considering the flare energy-frequency scaling for about 20 dMe
stars in the U and B photometric bands by \citet{gershbergShakhovskaya83}
and more recent studies of harder radiation flares summarized by
\citet{guedel03}, we estimate that the UV radiation environment of
very low-mass stars should be completely dominated by redistributed
flare energy.  The flares occur roughly once per day, with about
an order of magnitude variation in this rate.  Even the steady
coronal X-ray emission may be important for the most active of these
low-mass stars.

We expect biological activity and atmospheric chemistry to be
strongly influenced by the exposure to such intensely fluctuating
radiation environments, although the nature of the effects remains
to be estimated.  In particular, it is unknown whether such a
mutationally rich environment would enhance or suppress the rate
of evolution even in simple population genetics models.

\subsection{Stellar explosions}

Stellar explosions could also result in chemically and biologically
significant fluxes and fluences of ionizing radiation, albeit with
a much larger duty cycle than parent star flares.  Supernove produce
$\gamma$-ray emission associated with the radioactive decay of
freshly synthesized elements, mainly production of $^{56}$Ni that
decays to $^{56}$Co and then to $^{56}$Fe.  Monte Carlo calculations
of $\gamma$-ray deposition (e.g., \citealt{hoeflich98}) estimate a
Type Ia supernova release of \sci{6}{48} erg per Type Ia event (see
also \citealt{karam02}); Type II supernovae are much less important.
Using an average Galactic rate of Type Ia supernovae of \sci{3}{-14}
yr$^{-1}$ pc$^{-3}$ \citep{barbon99}, the average time between Type
Ia supernovae at distance $D_\mathrm{kpc}$ kiloparsecs is found to
be $T = \sci{8}{3} D_\mathrm{kpc}^{-3}$ yr.  If 1\% of the $\gamma$-rays
are redistributed to UV by the mechanism discussed in the present
work, we find that a biologically interesting fluence at the surface
of a planet should occur once every 10$^4$ yr.  However the associated
flux would be swamped by the parent star UV flux for a habitable
zone planet orbiting a solar-type parent star.  The redistributed
supernova UV flux will only exceed the parent star flux for low-mass
host stars that have a smaller fraction of their flux in the UV,
or for moons of giant planets at larger distances from solar-like
host stars.  We emphasize that the integrated mutation rate due to
SN explosions is negligible compared to the background mutation
rate because of the small durations compared to the recurrence
timescale.  Biologically, such intermittent hypermutation events
may be most important for partial sterilization of planets and
consequent effect on niche structure.

The $\gamma$-rays from supernovae can affect the atmospheric chemistry of
habitable planets of solar-type stars, independently of any UV
redistribution, through the direct effects of high-energy photons.
The chemistry resulting from irradiation of a present-day Earth
atmosphere was studied in detail by \citep{gehrels03} using a
single-scattering approximation for the radiative transfer.  The
more accurate transfer calculations in the present work agree fairly
well with their results for energy deposition as a function of
altitude in the thick-atmosphere, low-energy regime.  However the
expected rate of SN events near enough to significantly affect
atmospheric chemistry is estimated to be only 1-2 per Gyr
\citep{gehrels03}.

Supernova cosmic rays arriving later may be a more potent source
of shower $\gamma$-rays and fast particles.  From an evolutionary
perspective, such events are especially interesting because diffusive
propagation of cosmic rays implies long but uncertain exposure
durations from 100 yr \citep{ruderman74} to 10$^4$ yr \citep{shklovsky69}.
The modulation of Galactic cosmic rays by the astrosphere as planetary
systems pass through dense interstellar clouds (as suggested, for
example, by \citealt{begelmanRees76, zankFrisch99}) may be more
important than cosmic rays directly generated by the supernova
itself.  The statistics of fluctuations in astrospheric modulation
of cosmic rays are modeled in detail in \citet{smithScalo04}.

Finally, we consider gamma-ray bursts as a potential source of
intermittent ionizing radiation.  Their energy output is so large
that they could deliver a biologically important dose from essentially
anywhere in the Galaxy, although the duration, $\sim$ 10 sec on
average, is so small that the main effects would be either partial
sterilization of a planet or residual atmospheric chemistry
perturbations. Recent evidence favors strong redshift evolution of
the cosmic star-formation rate (e.g., \citealt{kewley04} and
references therein), which is needed to convert the observed gamma-ray
burst rate to a Milky Way rate, so we use the ``strong evolution''
rates in \citet{scaloWheeler02} to estimate the frequency of gamma-ray
bursts at a given fluence.  We find that the redistributed UV flux
will only exceed the solar UV flux about once per \sci{4}{8} yr,
with larger rates for lower-mass parent stars.  For ozone depletion,
the detailed study by \citet{gehrels03} of supernova direct gamma-ray
irradiation requires an above-atmosphere fluence of $\sim 10^8$ erg
cm$^{-2}$, giving a recurrence frequency of gamma-ray bursts of
this fluence of about 0.5 to 1 such events per Gyr, similar to but
a little lower than was found by \citet{gehrels03} for SN $\gamma$-rays.
The numbers are similar because the larger energies of gamma-ray
bursts are offset by the smaller rates per unit volume compared to
supernovae.

We conclude that the most important source of ionizing radiation
for both biological and chemical effects are flares from parent
stars, especially for low-mass stars.  Supernovae and $\gamma$-ray
bursts, because of their large duty cycle, are probably only important
if they induce partial extinction events, either directly through
lethal dose exposure, or by atmospheric chemistry alterations (e.g.,
\citealt{melott04}).  These events likely occur at a mean rate of
one per 0.1 to 2 Gyr, depending on the mass of the parent star and
the type of event involved.  By contrast, the ionizing radiation
environment of a habitable zone planet orbiting a low-mass star or
a young solar-mass star probably involves a steady and frequent
barrage of high-intensity flares with durations of minutes to hours
occurring at rates of once per week to once per hour, depending on
mass and age of parent star.  The present work shows that a significant
fraction of this radiation, rather than being absorbed high in the
atmosphere, can reach the stratosphere or the surface in the form
of redistributed UV radiation.

\subsection{Energetic particles}\label{sec:particles}

Although we are only explicitly concerned with the effects of photons
generated by astronomical events, high-energy particle emissions,
mainly solar energetic  particles (SEPs), solar chromospheric mass
ejections (CMEs), and Galactic cosmic rays are also of interest,
since their interaction with planetary atmospheres will  result in
the same kind of ultraviolet radiation through the generation of
secondary electrons along the primary particle path. Some comparison
of particles  with photons is afforded by the recent summary of
solar activity by \citet{smithMarsden03}. Their Fig.~3 shows the
flux of 1.8--3.8 MeV protons due to SEPs as a function of time
during solar maximum and minimum. At solar maximum, the flux is of
order 0.1 (cm$^2$ s sr MeV)$^{-1}$, with excursions up to two orders
of magnitude  larger and smaller. Since it is the total energy flux
that matters for redistribution to ultraviolet radiation, we convert
this to a total energy flux at  the mean SEP energy to obtain
\sci{1}{-5} erg cm$^{-2}$ s$^{-1}$. Even allowing for a two order
of magnitude enhancement, this flux is still small compared to a
10$^{32}$ erg solar flare at 1 AU, which is 10 erg
cm$^{-2}$  s$^{-1}$. The SEP flux is much more steady while the
flare flux is only intermittent, so the fluence from SEPs is larger;
however the redistributed UV flux from SEPs even at their peak is
negligible. On the young Sun or on more active, lower-luminosity
stars, the average flare energy and frequency is much larger, and
one might expect the energetic particle flux to keep in step (e.g.,
\citealt{wood02} on astrospheric momentum flux variations). We
arrive at similar conclusions by examining the data for CME $\sim$
6 MeV protons from the 14 Jul 2000  Bastille day event (data from
\texttt{http://soho.hascom.nasa.gov/hotshots} web  site). The energy
flux from Galactic cosmic rays is larger than that from solar cosmic
rays, at least for older stars like our Sun (see \citealt{smithMarsden03}),
and may vary considerably as the Sun travels through the Galaxy.
Therefore we cannot rule out the importance of Galactic cosmic rays
as a significant source of redistributed UV flux. 

One type of solar particle that does seem important is the so-called
``solar proton event,'' a short-duration SEP burst of $\gtrsim$ 10
MeV particles often associated with flares and presumably accelerated
by coronal mass ejection shocks.  The particle peak fluxes for the
35 most energetic of these events from 1973 to 2001 is given by
\citet{elborie03}.  The average flux of particles with energies
above 30 MeV is 0.2 erg cm$^{-2}$ s$^{-1}$, while the largest is
1.6 erg cm$^{-2}$ s$^{-1}$.  This is a lower limit because the
proton spectrum rises with decreasing energy down to at least 10
MeV, so these events rival the most energetic solar flares in energy
flux.  In fact \citet{sheaSmart96} list the strongest solar proton
event recorded as having a number fluence of \sci{3.4}{10} cm$^{-2}$
at 1 AU.  If the event lasted an hour (typical for SEP bursts), the
energy release at the Sun would be 10 times that of the largest
recorded solar flare. 

For any energetic particle flux, whether Galactic or solar in origin,
the resulting UV flux can be estimated by assuming most of the
cosmic-ray energy is deposited in 35 eV secondary electrons that
convert their kinetic energy into UV auroral radiation with the
same efficiencies as found in the present paper for secondary
electrons resulting from photon events.

\section{Summary and Conclusions}

The continuum UV emission from the Sun would have been very
intense during the Archean era before the development of the ozone
layer. Furthermore, the Sun was likely to have been much more
active in the past when life first gained a foothold on Earth.
\citet{guinanRibas02} show that the coronal and X-ray to extreme
UV emission of the young Sun were 100--1000 times stronger than
those of the present Sun.  Even now, solar flares are significant:
they follow a power law fluence-per-interval relation that
suggests that more powerful, but less frequent, flares are likely
even for quiescent, aging solar type stars \citep{aschwanden00}.
Mars may have once had a thick atmosphere that would still be
subject to the strong redistribution of ionizing radiation into
auroral UV in the manner we describe here, and it is now very
susceptible to direct incident irradiation. Expanding our
perspective to other stars hosting other planets, the case can
easily be made that UV and ionizing radiation, including
stochastic bursts of hard radiation, are the norm in our
tumultuous Galaxy \citep{scalo03}.

To establish quantitatively the effects of ionizing radiation in
terrestrial-like exoplanet atmospheres, we have used Monte Carlo
models to propagate ionizing radiation through a suite of simple
model atmospheres. We constrained the parameter space of the
atmospheres by limiting the models to conditions that are
consistent with ``habitable'' planets, in the sense that the
atmosphere is thick enough to maintain liquid water on the
surface, given enough ambient heating to keep the water in liquid
form.  We estimate the lower limit for atmospheric column depth
for habitable planets to be about 30 g cm$^{-2}$.  Above this
limit, we characterize two types of atmospheres: ``thin'' and
``thick.''

Our results can be summarized as follows.

\begin{enumerate}

\item Thin atmospheres with column density less than about 100 g
cm$^{-2}$ will directly transmit a substantial portion of any
incident $\gamma$-ray flux.  Even for these thin atmospheres,
incident X-rays will be blocked because of the high cross section
for photoabsorption.  Contemporary Mars represents an example of
this sort of thin atmosphere.

\item For planets with relatively thin
atmospheres, the ionizing radiation spectrum at the surface from
solar flares, supernovae, gamma-ray bursts or other sources of
hard radiation should be relatively flat above 50--100 keV due to
Comptonization, with a low-energy, photoabsorption cutoff.

\item We define thick atmospheres to be those with column density
in excess of about 100 g cm$^{-2}$, in which both $\gamma$-rays
and X-rays will be blocked.  In this case, however, we show that,
in the absence of UV blocking agents (O$_3$ or aerosols for
instance), a substantial fraction of the incident energy will
still arrive at the planetary surface as UV resulting from
molecular excitation by secondary electrons produced by the
Compton scattering of primary radiation and associated primary
photoionization electrons. Typically 1--10\% of the incident
energy can reach the ground as this biologically-active
``auroral'' UV.  This condition is typical of the Archean Earth
where the only opacity to UV may be Rayleigh scattering.

\item A significant fraction of the incident energy may reach the
surface even for contemporary Earth with its O$_3$ shield.  We
estimate than even today, a fraction of order \sci{2}{-3} of
incident hard flux will reach the surface of the Earth in the form
of UV radiation in the 200--320 nm band, independent of the form
of the incident ionizing radiation spectrum.

\item The spectrum of the redistributed UV radiation arriving at
the planetary surface will depend on the rich and complex
molecular emission line spectrum.  We have considered relevant
bands of N$_2$ to estimate that the net effect can be approximated
by a continuous spectrum in which the energy flux is distributed
approximately inversely with wavelength. We argue that essentially
all molecules that might be substantially represented in the
atmosphere of a habitable exoplanet would have electronic levels
with similar spacings that would be excited with an efficiency
comparable to N$_2$.

\item We show that the results are not substantially affected by
thermalization of the incident radiation since the ionized
fractions of the atmosphere are typically low, nor by quenching,
i.e., collisional deexcitation of the molecules, at the typical
low electron densities, nor by the angle of incidence.

\item Our results show that low altitude ``secondary ionospheres''
can be produced in thick atmospheres if the ionizing radiation
source is a supernova or gamma-ray burst. In the case of
stellar flares, the existing ionospheres of thick atmospheres will
be further ionized by a substantial margin.  Ionization fractions
in all cases of irradiation that exceed the parent star continuum
are comparable to or greater than the steady-state terrestrial
ionospheres. This phenomenon could affect atmospheric chemistry
and global climate, especially in the case of the more frequent
stellar flares on low-mass stars.

\item We do not know if early Mars had a thick atmosphere and, if
so, whether or not it contained UV blocking agents
\citep{haberle94,leblancJohnson01}.  The present work suggests
that, even if it did, its early evolution, when life might have
been forming or expanding through evolutionary niches, would have
been punctuated by bursts of reprocessed UV from stellar flares at
relatively frequent intervals and again by more exotic but
inevitable astronomical events at larger intervals.

\end{enumerate}

Planets with thick atmospheres can be shielded from direct
ionizing radiation and even from ordinary continuum UV if their
atmospheres contain effective UV shields and still be subject to
bursts of biologically significant UV. \citet{smith04oleb} estimate
that steady-state solar UV could be exceeded by redistributed UV
from intense solar flares roughly once per decade.  The
redistributed flare energy rapidly increases in importance for the
very common lower mass stars that have less continuum UV flux and
more intense and frequent flares (see \citealt{guedel03}).

The point of view that much of terrestrial and extraterrestrial
life is driven by radiation sources was first outlined in the
classic book by \citet{shklovskiiSagan66}, but has lain
substantially dormant since. \citet{rothschild99} discusses a
large number of possible relations between radiation and
biological evolution. Significant aspects of evolution itself may
be in response to changing radiation environments.  Much DNA
damage is either not repaired, leading to cell death, or is
repaired precisely, in which case there is no mutation.  In
neither extreme is there evolution. On the contrary, a significant
amount of current-day mutation is due to error-prone light-induced
DNA damage repair of cyclobutane pyrimidine dimers incurred by UV
radiation (e.g., \citealt{alpen98,jagger85,vonsonntag87}). In
addition, the mechanisms involved in the repair of DNA damage due
to UV and ionizing radiation are often the same as those involved
in gene transfer and meiosis (\citealt{michodWojciechowski94} and
references therein). It is conceivable that early life had to
learn the techniques of radiation repair for survival, but then
adapted them to powerful modes of evolution, first lateral gene
transfer and then sexual reproduction. In this context of the
possible fundamental importance of UV damage and repair, it is
then especially interesting that planets with thick atmospheres
that will shield surficial life from direct ionizing radiation
will nevertheless shower the surface with UV irradiation in
response to the stochastic astronomical radiation environment from
the host star and more distant, yet significant Galactic events.

\section*{Acknowledgments}

We thank Jim Kasting and Alex Pavlov for pointing out the two-stream
solution for Rayleigh scattering, David Lambert for lending us
nitrogen energy level data, Peter H\"oflich for first suggesting
that the redistributed flux could be large, and two anonymous
referees for comments and suggestions.  DSS thanks the Harrington
Doctoral Fellows Program and the NSF Graduate Research Fellowship
Program for support. In addition, we gratefully acknowledge support
from NSF grants AST 9907582 and AST 0098644.

\appendix

\section{Collisional Quenching}\label{sec:quench}

In our model, we assume all photon energy deposited in the
atmosphere is reemitted as UV because of the efficiency of
secondary electron excitation in a gas of very low ionization
fraction. But in reality part of the reemission will be quenched
by collisional deexcitation. Quenching was not included in our
calculation because it would require solving the complete non-LTE
level population rate equations for a variety of potential
atmospheric constituents, a level of complexity and uncertainty
beyond the scope of the present work. Nevertheless, we do wish to
estimate its importance.

We first consider the usual two-level approximation. Rigorously,
the two-level solution for the line intensity cannot be used because
the principle of detailed balance between excitation and deexcitation
rates does not hold when the secondary electrons have a non-Maxwellian
velocity distribution. We instead require that all electron excitations
result in an emitted line photon, except for the fraction suffering
collisional deexcitation.

Ionization fractions are small enough in terrestrial-like exoplanet
atmospheres that deexcitation occurs primarily via neutral atoms
and molecules.  The exception is the highest altitudes of atmospheres
subjected to very high fluence ($\gtrsim 10^8$ erg cm$^{-2}$) stellar
flares.  We assume that the quenched transition is not forbidden,
which would reduce the Einstein $A_{ji}$ value by a large factor,
as in terrestrial [OI] emission.  The following method
also applies (with some modification) for vibrational transitions
within a given electronic level.

For the following estimate, we compute the excitation-deexcitation
balance and obtain the relative importance of collisional
quenching and radiative deexcitation in the most important N$_2$
auroral emission band systems listed in Table
\ref{table:auroralLines}. The N$_2^+$ level is important
despite low ionization fractions because many of the secondary
electrons will ionize N$_2$ to the B$^2\Sigma_u^+$ excited of
state of N$_2^+$, and subsequent fluorescence to the
X$^2\Sigma_g^+$ state yields the well-known strong 391.4 nm
auroral band. The cross section for this process is large, and the
efficiency of production of this band relative to all ionizations
is about 6\% (\citealt{banksKockarts73}, p.~213). Following
\citet{jones74}, we let $dn_2/dt$ be the rate of excitation of the
target molecule---N$_2$ in our case---to the upper electronic level
by secondary electrons in the two-level scheme, $A_{21}$ is the
Einstein A value for the downward transition, $Q_{21}$ is the
thermally averaged collisional rate coefficient $\langle \sigma v
\rangle$ for downward transitions due to collisions between the
target and the dominant quenching particle M (N$_2$ and O$_2$ for
the Earth), and $n_2$ and $n_\mathrm{M}$ are the number densities
of the excited species and quenching species M, respectively. The
balance between secondary electron excitation and the sum of
radiative and collisional deexcitation can be written as \beq
\frac{dn_2}{dt} = A_{21} n_2 + Q_{21} n_2 n_\mathrm{M}.\eeq  By
dividing the balance equation by the unquenched rate $A_{21} n_2$,
it is easy to show that the unquenched radiative deexcitation rate
is reduced by a quenching factor $f_Q$: \beq f_Q= 1+ n_\mathrm{M}
Q_{21} /A_{21}.\eeq We also define the critical density of
quenching particles to be $n_\mathrm{M,crit}\equiv A_{21}/Q_{21}$,
at which the emission is halved.

\begin{table} \begin{center} \begin{tabular}{lccc}
\hline
Band &
$A_{ji}$ [s$^{-1}$]& $Q_{ji}$[cm$^3$ s$^{-1}$] &
$n_\mathrm{M,crit}$ [cm$^{-3}$]\\
\hline
Vegard-Kaplan & 0.53   &
\sci{1.5}{-11} & \sci{3}{10}     \\
L-B-H         & \sci{8.3}{3} &
$\lesssim \sci{3}{-10}$ & $\gtrsim \sci{3}{13}$  \\
Herman-Kaplan & \sci{5.3}{3}
& $\sim 10^{-10}$ & $\sim \sci{5}{13}$ \\
2nd positive  &
\sci{2.7}{7} & $\sim 10^{-10}$ & $\sim \sci{7}{17}$ \\
1st
negative & \sci{1}{7}   & \sci{4}{-10}   & \sci{3}{16}   \\
\hline
\end{tabular}
\caption{\label{table:quenchData}Quenching factor data for N$_2$
UV band systems
\citep{banksKockarts73,lofthusKrupenie77,huberHerzberg79}. See
Table \ref{table:auroralLines} for definitions of the bands.}
\end{center}
\end{table}

We have estimated the critical quenching height $z_Q$ in our
models, at which $n_\mathrm{M}(z_Q)=n_\mathrm{M,crit}$, and the
altitude of maximum energy deposition $z_E$, for two important UV
transitions of N$_2$. We can then gauge the amount by which a line
is quenched by defining the quenching ratio, $\rho_Q\equiv
z_Q/z_E$.  When $\rho_Q \ll 1$, excitations take place where
densities are low enough that collisional deexcitation is
unimportant.  The results of this approach can be applied to any
other molecule of interest, depending on $A_{ji}$ and $Q_{ji}$.
Table \ref{table:quenchData} lists the relevant parameters for the
N$_2$ band systems of Table \ref{table:auroralLines} taken from
\citet{lofthusKrupenie77}, \citet{huberHerzberg79}, and
\citet{banksKockarts73}. Since the lifetimes of the upper
molecular electronic states vary by orders of magnitude, we have
chosen to illustrate the situation with two representative
transitions of N$_2$ in Table \ref{table:quenchData}---the
Lyman-Birge-Hopfield and 2nd positive systems.  These bands have
$A_{ji}$ values (10$^4$ and 10$^8$, respectively) that cover the
range of values for allowed transitions.

\begin{figure} \centerline{ \plotone{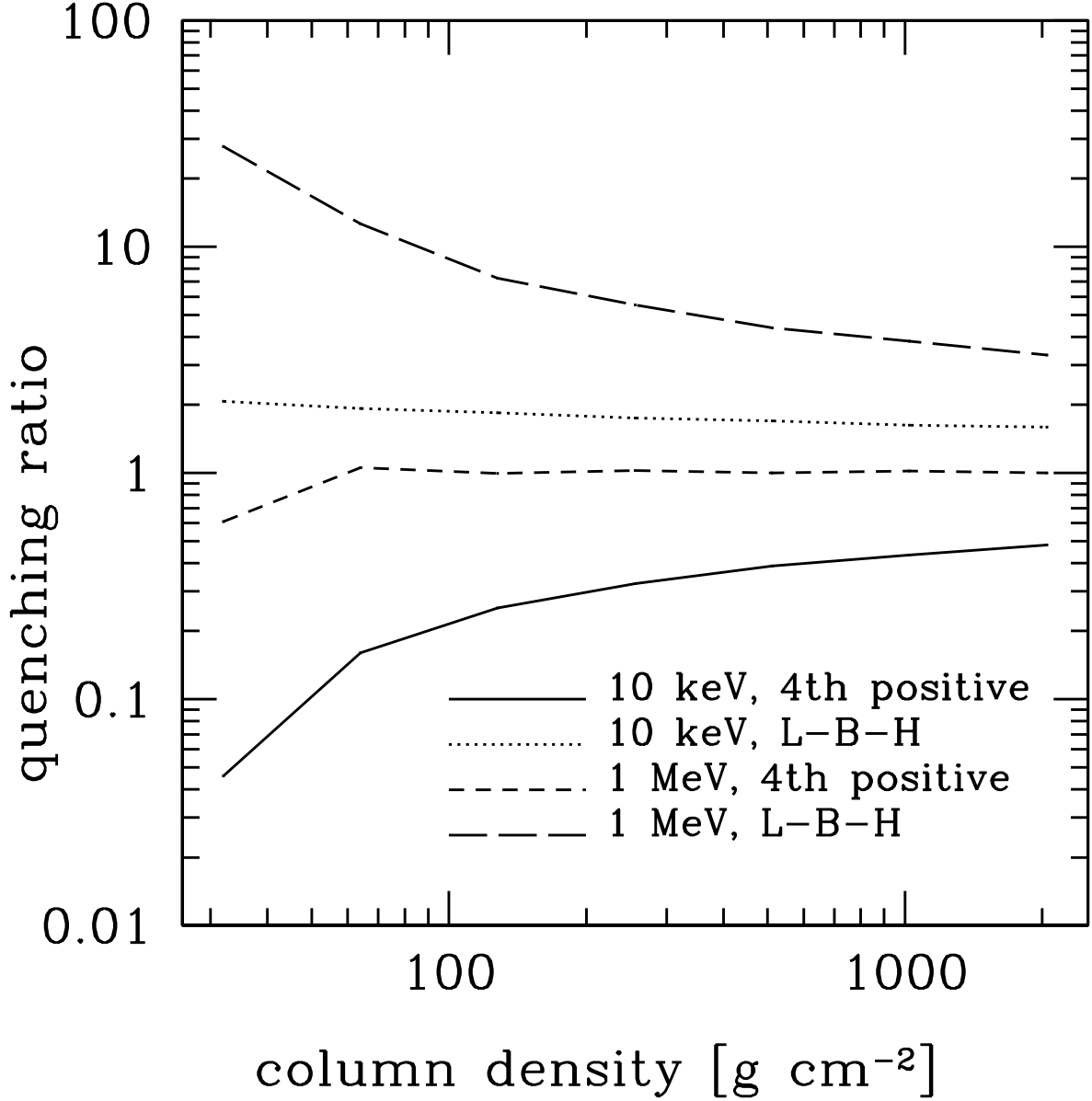}
} \caption{\label{fig:quenchRatioCD}Quenching ratio $\rho_Q$ as a
function of column density for two representative molecular
nitrogen systems. Only the Lyman-Birge-Hopfield system is
significantly quenched and only at very high incident energies,
for which the altitude of energy deposition $z_E$ is very low. For
hard X-ray incident energies, neither of the systems is completely
quenched, but the reemission in the L-B-H band would be reduced by
a factor of a few. The quenching ratio approaches unity for very
thick atmospheres (see text for explanation).}
\end{figure}

Figure \ref{fig:quenchRatioCD} shows $\rho_Q$ as a function of
atmospheric column density for the two representative N$_2$ systems
and two different incident photon energies.  We can see from the
plot that only Lyman-Birge-Hopfield (small $A_{ji}$ value) is
significantly quenched and only at very high incident energies, for
which the altitude of maximum energy deposition $z_E$ is very low.
For X-ray incident energies, neither of the lines is significantly
quenched, but the reemission in the L-B-H band would be reduced by
a factor of about two.  We can then conclude that reemission due
to stellar flares incident on thin atmospheres are the least quenched,
while the highest-energy irradiation by supernovae and gamma-ray
burst $\gamma$-ray lines will be the most quenched.

\begin{figure} \centerline{
\plotone{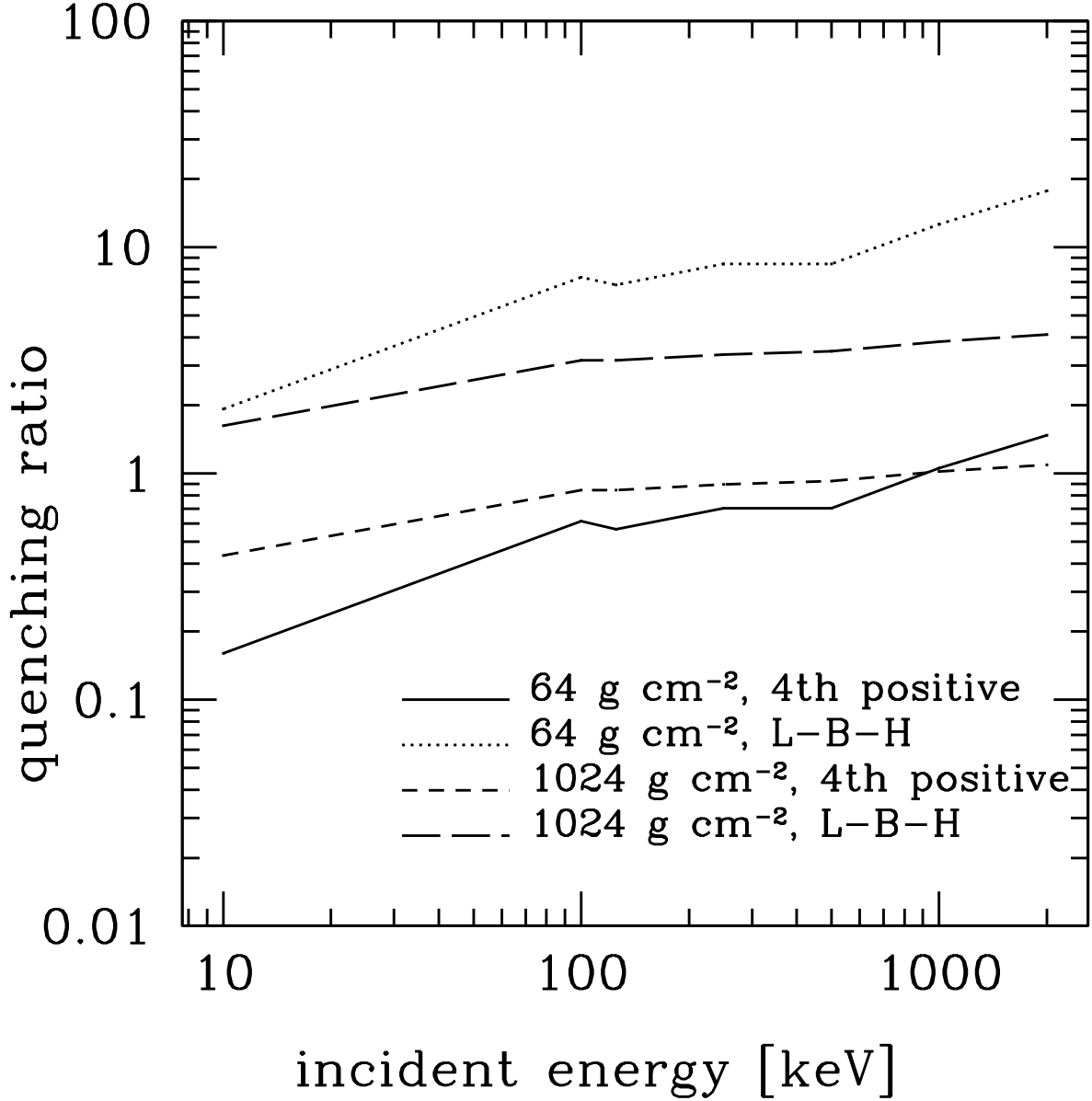} }
\caption{\label{fig:quenchRatioEnergy}Quenching ratio $\rho_Q$ as
a function of incident energy for the two representative N$_2$
lines. Again we see that the L-B-H is quenched more than the 2nd
positive transition, due to its lower $A_{ji}$.  Interestingly, we
again see that the difference in $\rho_Q$ for the two lines is
smaller for higher column densities, as can also be seen in
Fig.~\ref{fig:quenchRatioCD} (see text for an explanation).
Irregularities in the 64 g cm$^{-2}$ curve are due to the smaller
optical depth, i.e., fluctuations in the optical depth of the
height of maximum energy deposition are magnified in altitude,
leading to a larger uncertainty in $z_E$.}
\end{figure}

The quenching effect as a function of incident energy is shown in
Fig.~\ref{fig:quenchRatioEnergy}.  As expected the magnitude of
quenching increases with increasing incident energy, since $z_E$
decreases roughly logarithmically with optical depth.  Again we
see that the L-B-H system is quenched more than the 2nd positive
transition, due to its lower $A_{ji}$.  Interestingly, we see in
both Figs.~\ref{fig:quenchRatioCD} and \ref{fig:quenchRatioEnergy}
that the incident energy sensitivity of $\rho_Q$ is smaller for
higher column density atmospheres. This behavior can be understood
from the dependence of the quenching ratio on the optical depth
(which is proportional to the column density).  Using the Chapman
solution for the energy deposition, we can approximate the height
of maximum as $z_\mathrm{max}=h\log\tau$ (see
\S\ref{sec:chapman}). We defined the altitude at which quenching
becomes significant as $z_Q$ such that $n(z_Q)=n_\mathrm{M,crit}$.
From this we have, \beq z_Q=-h\log(n_\mathrm{M,crit}/n_0),\eeq
where $n_0$ is the number density of quenching molecules at the
planet's surface and $h$ is the scale height.  Taking the
definition of the quenching ratio, we can write an approximation
for it as \beqa
\label{eq:quenchApprox}\rho_Q &=& z_Q/z_E \nonumber\\
&=&-\frac{h\log(n_\mathrm{M,crit}/n_0)}{h\log\tau} \nonumber\\
&=&\frac{\log(Q_{ji}\Sigma/A_{ji}h)}{\log(\Sigma\sigma)},\eeqa
where $\Sigma$ is the column density and $\sigma$ is the cross
section for energy deposition at the original incident energy.  We
can see from the form of this formula that the ratio
$Q_{ji}\Sigma/A_{ji}h$ determines whether the quenching ratio is
smaller or larger than 1, since all atmospheres of exoplanets
considered habitable in this work have optical depths greater than
unity at the incident energy. The only energy dependence enters in
the denominator, in the optical depth $\tau$. Rewriting, \beq
\rho_Q = 1 + \frac{\log(Q_{ji}/A_{ji}\sigma h)}{\log\tau}.\eeq Now
we can see that as the column density (and hence $\tau$)
increases, the quenching ratio will approach unity. Furthermore,
the quenching ratio increases as the incident energy increases
because the dominant cross sections at keV to MeV energies
(photoabsorption and Compton scattering) both decrease with higher
energy.  In other words, higher penetration of the atmosphere as
the energy of the incident ionizing radiation increases reduces
the UV reemission efficiency by  depositing more of the energy in
denser regions of the atmosphere.

Depending on the relative fluxes of the various auroral lines, the
overall UV reemission will be quenched by a factor somewhere
between the limits given in the plots.  Since characteristic
$A_{ji}$ values and collisional deexcitation cross sections show a
similar range for other molecules we can generalize our
conclusions to the statement that quenching will only
significantly affect the surficial fluences for transitions with
$A_{ji}$ values of $\lesssim 10^4$ s$^{-1}$.  Given that all of
the lines listed in Table \ref{table:auroralLines} are roughly
equally strong, we expect that quenching will be insignificant for
stellar flare irradiation, and for supernovae and gamma-ray bursts, the
emission will likely be reduced by factor of only a few, depending
on the $A_{ji}$ value and column density.

\section{Weighted Monte Carlo transport algorithm}
\label{sec:mcalg}

An outline of the structure of the Monte Carlo code is as follows:

\begin{enumerate}

\item \textbf{Initialize}

\begin{enumerate}

\item Assign each photon an initial incident energy $E_0$ drawn
randomly from the specified incident spectrum.

\item Initialize the direction of propagation, $\theta_p$. The
angle of incidence, $\theta_i$, is defined with respect to the
plane of the atmosphere, but $\theta_p$ is oriented toward the
ground. Hence for a normally incident ray $\theta_p=0$ and
$\theta_i = \pi/2$. We track the direction of propagation in three
dimensions by a unit vector $(v_x,v_y,v_z)$ that represents the
direction of travel with respect to the ground, where the z-axis
points downward to the ground.  For the physics in this work,
however, only the projection of the vector onto the z-axis ($v_z$)
matters, where $v_z = \cos\theta_p$.

\item Set the statistical weight, $w$, of each photon to unity.
Higher weights imply that the photon represents a packet of $w$
photons, rather than one photon, which alters only normalizations.

\end{enumerate}

\item \textbf{Propagate}

\begin{enumerate}

\item Recalculate the total cross sections for absorption and
scattering.

\item For downward traveling photons, add a portion of the photon
energy equal to $w\exp(-\tau/v_z)$ to the spectrum of flux received
at the ground, where $w$ is the current weight of that photon,
$\tau$ is the optical depth from the photon's current altitude to
the ground, and $v_z$ is the downward component of the direction
vector from above.  This is the forced scattering procedure
(cf.~\citealt{witt77}).  The fraction of the weight removed that
represents the probability that the photon is still scattering
(which we are forcing it to do) is simply one minus the probability
that it did not scatter.  If the photon is directed upward, the
procedure is identical except the peeled-off weight is added to the
spectrum of photons reflected by the planet and $\tau$ corresponds
to the optical depth from the photon's altitude to the top of the
atmosphere.

\item Sample a random optical depth to the interaction location
from an exponential probability distribution $p(\tau)=\exp(-\tau)$
by generating a uniform deviate $R \in [0,1]$  and inverting $p$
to find the corresponding optical depth: $\tau = -\log R$. We choose
an exponential distribution with unit mean because the photon behaves
as part of a beam subjected to extinction and so has an intensity
following Beer's Law, or $I(\tau)=I_0 \exp(-\tau)$. Thus the
probability that a photon will traverse a distance corresponding
to an optical depth $\tau$ unimpeded is $\exp(-\tau)$.

\item Move the photon a distance in the atmosphere corresponding
to the randomly sampled optical depth.  Since the atmosphere is
based on an exponential density profile, the new altitude can be
found analytically, eliminating one of the most computationally
intensive procedures in Monte Carlo radiative transport---sampling
a density field along a ray.  From the sampled optical depth
$\tau$, we update the altitude $z$ to $z'$ according to \beq
 z' = -h \log\left[ \exp\left(-\frac{z}{h}\right) +
 \frac{v_z\tau}{\chi h}\right], \eeq where $\chi$
is the total extinction coefficient, including Compton scattering
and photoabsorption, and $h$ is the scale height.

\end{enumerate}

\item \textbf{Interact}

\begin{enumerate}

\item At the new location $z'$, multiply the statistical weight
$w$ by the scattering albedo $a$, which represents the probability
that the photon still exists after the interaction. (If the photon
were instead replaced by a continuous energy stream, a fraction
$a$ of the flux that interacted would be scattered, while a
fraction $1-a$ would be absorbed.)

\item Add the fraction of energy that was photoabsorbed to the
spectrum of energy deposited at this layer. This represents the
fraction of photons that would not have been scattered.  Rather
than absorbing all the energy of a fraction $1-a$ of a packet of
photons, the weighting technique stipulates instead to absorb that
fraction of the energy of a single photon.

\item Choose a new propagation direction $\theta_p$ by sampling by
a rejection technique from the differential Klein-Nishina formula,
where the forward direction is parallel to $(v_x,v_y,v_z)$.  The
angular distribution is symmetric about this direction, so the
angle about that axis is chosen from a uniform distribution.  The
sampled angle gives the direction change after the scattering
event, so the a new direction vector must be calculated.

\item Update the photon energy based on the change in direction
and the corresponding Compton energy loss. Add the Compton recoil
electron energy to the spectrum of energy deposited in this layer.
This energy is multiplied by the scattering albedo and the current
photon weight in order to conserve energy.

\item Remove the current photon from the model if the updated
weight is smaller than a predetermined minimum value, since the
photon is now statistically insignificant; otherwise, go to step 2
and repeat.

\end{enumerate}

\end{enumerate}

\section{Code benchmarks}\label{sec:codebench}

\subsection{Pure scattering atmosphere}
\label{sec:scattTest}

A standard solution to the radiative transfer equation can be
obtained for pure scattering in the two-stream approximation.  The
approximation of pure scattering given by \citet{schuster05} assumes (i) a
plane-parallel atmosphere, (ii) a diffuse source incident at the
top of the atmosphere, (iii) no sources inside the atmosphere, and
(iv) a perfectly absorbing ground.  With these conditions, the fraction of
the incident flux transmitted through the atmosphere is $T\simeq
2/(2+\tau/\mu)$, where $\tau/\mu$ is the total optical thickness
of the atmosphere divided by an angle cosine which represents the
average angle of incidence of the radiation field. The actual
value of $\mu$ can only be obtained by iteratively solving the
radiation field until a value for $\mu$ converges; however as can
be seen in our benchmarks, the approximation $\mu\simeq 1$ for
normally incidence radiation is sufficiently accurate.

\begin{figure}
\centerline{ \plotone{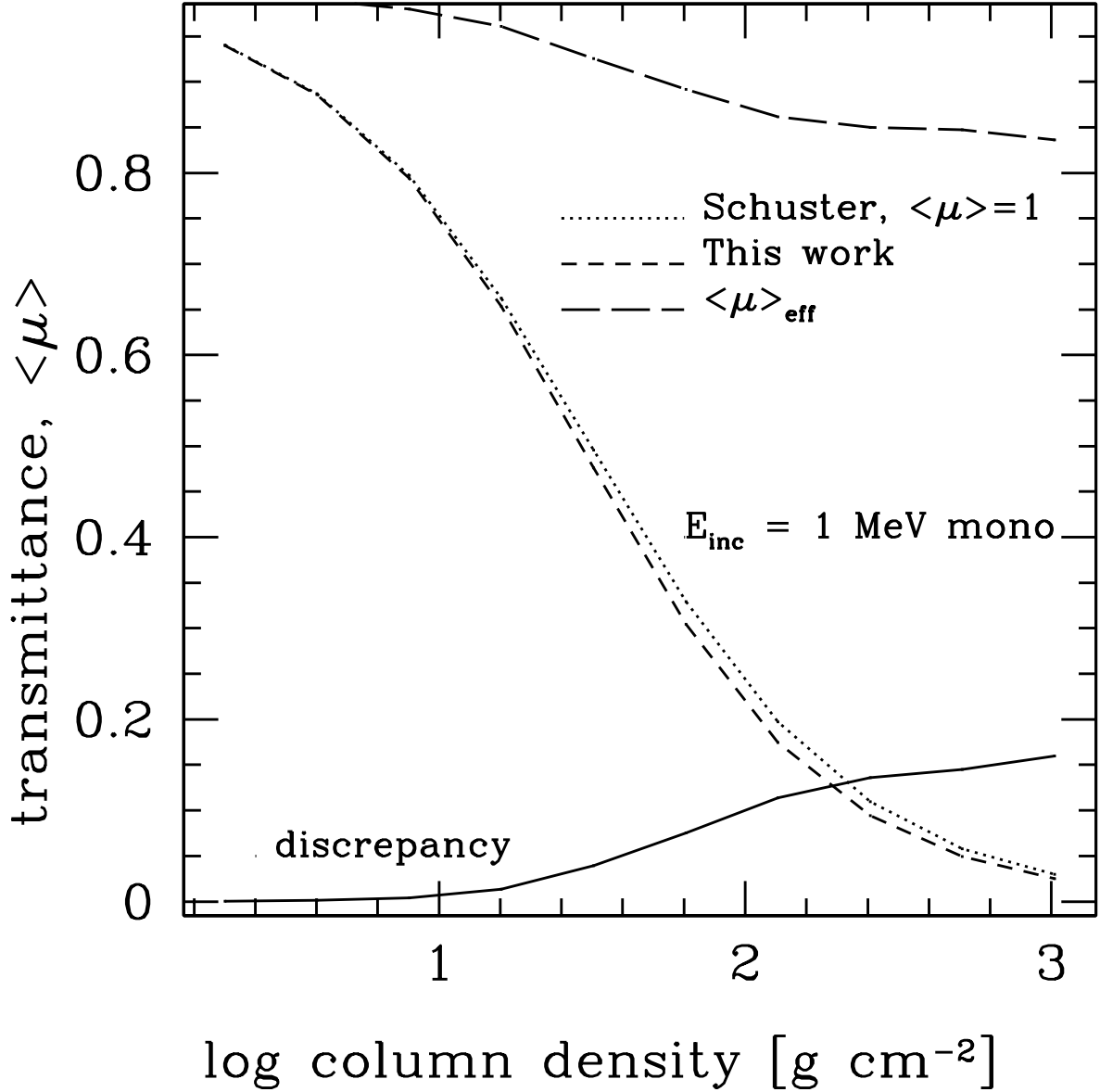} }
\caption{\label{fig:scattTest}Comparison of the Monte Carlo code
with the Schuster (1905) pure scattering solution. The agreement
is quite good, even for thick atmospheres, where the radiation
field deviates most from being monodirectional.  The fractional
discrepancy between the two results is due to the approximation of
$\avg{\mu} = 1$ for the Schuster result, which would tend to
overestimate the transmittance. The value of
$\avg{\mu}_\mathrm{eff}$ required to bring the Schuster data down
to ours is shown and is $\gtrsim 0.8$ even for the thick
atmospheres, implying that $\avg{\mu} \sim 1$ is an acceptable
approximation. In our calculations, we assume normal incidence.}
\end{figure}

To compare our code, which includes more complicated physics than
 pure scattering, to a known scattering solution, we removed
photoabsorption and altered the treatment of Compton scattering to
make it conservative. Thus the photons were allowed to scatter
with a cross section equal to the Compton cross section, but the
energy changes were ignored.  Figure \ref{fig:scattTest} shows the
comparison between the Monte Carlo code and the Schuster solution as a
function of column density. The agreement is quite good for a
Monte Carlo code, even for thick atmospheres, where the
discrepancy is $\lesssim 15$\%. For reference, the $\mu$ required
to bring the Schuster transmittance into agreement with ours is
shown.

\subsection{Beer-Lambert absorption}

In the case of pure absorption, photons interact only once and then
are removed from the model.  To simplify the calculation, we treated
the extinction coefficient as purely absorptive and removed all
photons upon the first scattering.   We found the agreement between
the exponential attenuation approximation of the Beer's law and
the Monte Carlo code to be excellent.  The Monte Carlo code shows
an exponential dependence and matches Beer's Law to better than 1
part in $10^4$ for even the thickest atmospheres.  This test is
somewhat trivial, but disagreement would nevertheless indicate
problems with the radiative transfer code.

\subsection{Comptonization by cold electrons}

A third test was performed to test solely the non-conservative,
Compton scattering aspect of the code.  We removed all
photoelectric absorption and allowed each photon to Compton
scatter a fixed number of times (100, in this case).  In the limit
of large scattering number, the Compton energy losses become
small, and the photon energy spectrum approaches a Gaussian. Based
on the results of \citet{xu91}, we can write an analytic
approximation for the spectrum as a function of initial energy and
scattering number. The energy spectrum after the $n$-th scattering
is given by Eq.~10 of \citet{xu91}: \beq\label{eq:xu} F_n(\lambda) = (2\pi
\sigma_n^2)^{-1/2} \exp[-(\lambda-\lambda_n)^2/2\sigma^2_n], \eeq
where \beq \lambda_n = \lambda_{n-1} + 1 -
\frac{4}{5\lambda_{n-1}} +
O\left(\frac{1}{\lambda_{n-1}^2}\right),\eeq \beq \sigma_n^2 =
\left(1+\frac{8}{5\lambda_{n-1}^2}\right)\sigma_{n-1}^2 +
\frac{2}{5} + O\left(\frac{1}{\lambda_{n-1}^2}\right),\eeq and
$\lambda$ is in units of the Compton wavelength ($\lambda_c \equiv
h / m_e c$).

\begin{figure}
\centerline{ \plotone{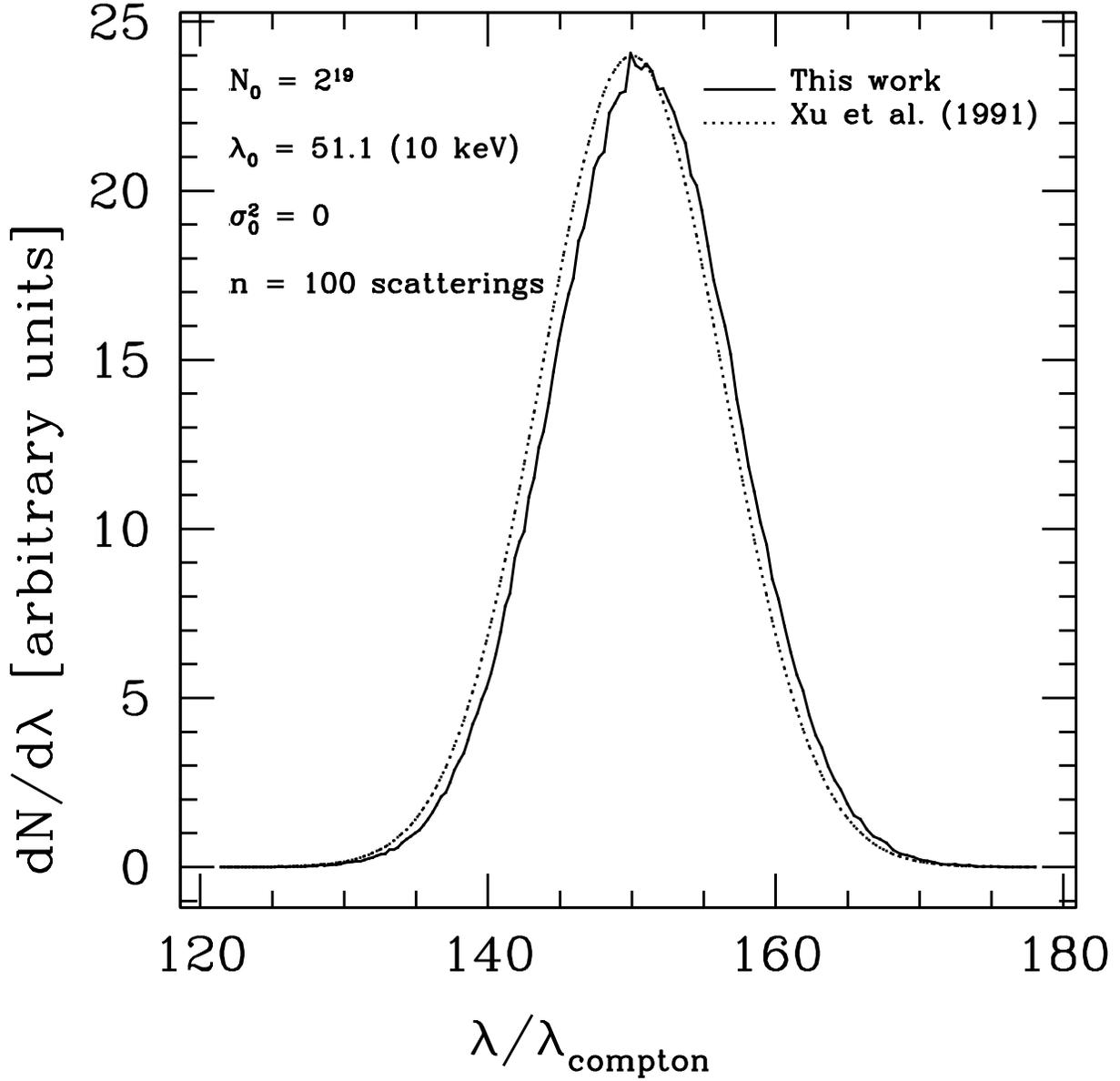} } \caption{\label{fig:xuTest}
Comparison of the Monte Carlo code with the results of \citet{xu91}. The
slight shift to longer wavelengths of our code compared to the
analytic approximation is due to the neglect of the higher order
terms in the analytic expression for $\lambda_n$ and $\sigma_n^2$ (see text),
which leads to an underestimation of the peak wavelength and
variance when using the \citeauthor{xu91} formula (Eq.~\ref{eq:xu}).}
\end{figure}

Figure \ref{fig:xuTest} shows a comparison of the Monte Carlo code and the Xu
et al. formula for $n=100$ scatterings of $2^{19}$ photons, each
with an initial dimensionless wavelength ($\lambda/\lambda_c$) of
51.1 (equivalent to an energy of 10 keV). The correspondence is
excellent. The slight shift to longer wavelengths of the Monte Carlo results
compared to the analytic approximation is due to the neglect of
the higher order terms in the above formula for $\lambda_n$ and
$\sigma_n^2$, which leads to an underestimation of the peak
wavelength and variance when using the \citeauthor{xu91} formula.
The Monte Carlo code uses the full Compton energy shift formula and energy
dependent cross section and thus should be more accurate.

\section{Modification to the Schuster solution}
\label{sec:schusterModification}

\citet{schuster05} solved the problem of a pure scattering atmosphere
for a source above the atmosphere and a purely absorptive base (see
\S\ref{sec:scattTest}).  For the problem of the transmission of UV
reemission in a planetary atmosphere, we adopt an average incidence
angle cosine $\avg{\mu} = 0.5$, since our source of reemission is
isotropic, and we expect the radiation field to be roughly so.
Under these boundary conditions the fraction of the incident flux
transmitted through the atmosphere is \beq T(\tau) = \frac{1}{1+\tau},
\eeq where $\tau$ is the total optical depth of the atmosphere.
Similarly, the albedo of the atmosphere is $R \equiv 1 - T$, or
\beq R(\tau) = \frac{\tau}{1+\tau}.\eeq

\begin{figure}
\centerline{ \plotone{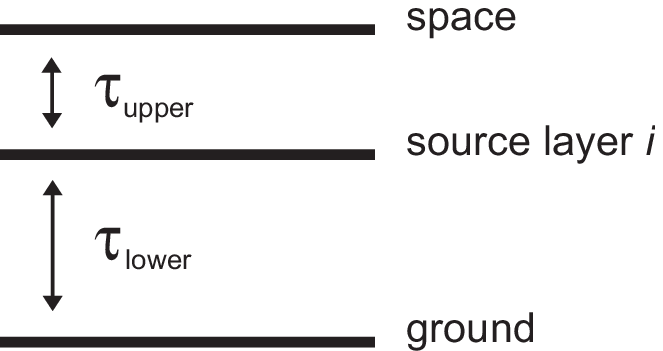} }
\caption{\label{fig:schusterDiag}Diagram of the geometry of the UV
redistribution layer and the surrounding atmosphere.}
\end{figure}

For the UV redistribution in our work, we modified this solution
to accurately handle reflection from the part of the atmosphere
above each UV reemission layer, which was not present in the
original Schuster solution. In our case, on a layer by layer
basis, we have an isotropic source (a layer at which
redistribution from ionizing to UV radiation occurs) sandwiched
between two atmospheres with purely absorptive boundaries (both
the ground and space do not reflect).  The situation is
illustrated in Fig.~\ref{fig:schusterDiag}. We denote the optical
depths of the upper and lower atmospheres as $\tau_\uparrow$ and
$\tau_\downarrow$, respectively.  Since we are concerned with the
amount of reemitted UV which reaches the surface, we set the
transmission fraction of the ``sandwich'' to be the sum of the
flux directly transmitted from the emission layer to the ground
and the flux reflected between the two atmosphere layers and
finally transmitted to the ground.  Each atmosphere layer obeys
the Schuster solution in isolation, but together the reflection
terms increase the expected transmission by a significant margin.
In a manner similar to the popular two-stream approximation, the
isotropic source flux can be divided into a downward hemisphere
$(\mu_+ \ge 0)$ and an upward hemisphere $(\mu_- < 0)$, each
containing half of the total emitted flux.  Starting with the
downward hemisphere, the successive contributions to the surface
flux from multiple reflections can be written easily.  Starting
with the flux transmitted without reflection, we add the
contribution from the flux that has reflected once off the bottom
atmospheric layer and then off the top atmospheric layer and is
then transmitted through the bottom atmospheric layer. To that we
add the flux that reflects off the bottom atmospheric layer, then
the top, then the bottom, then the top, and then is transmitted
through the bottom atmospheric layer, etc.  Hence, \beqa T(\mu_+)
&=& \frac{1}{1+\tau_\downarrow} +
\frac{\tau_\downarrow}{1+\tau_\downarrow}
\frac{\tau_\uparrow}{1+\tau_\uparrow}\frac{1}{1+\tau_\downarrow}
\nonumber\\ & & + \left(\frac{\tau_\downarrow}{1+\tau_\downarrow}
\right)^2 \left(\frac{\tau_\uparrow}{1+\tau_\uparrow} \right)^2
\frac{1}{1+\tau_\downarrow} \nonumber\\ & & + \cdots \nonumber\\ &
= & \frac{1}{1+\tau_\downarrow} \sum_{p=0}^\infty \left[
\frac{\tau_\downarrow\tau_\uparrow}{(1+\tau_\downarrow)(1+\tau_\uparrow)}
\right]^p. \eeqa Since $\forall \tau_\downarrow, \tau_\uparrow >
0$, we have \beq
\frac{\tau_\downarrow\tau_\uparrow}{(1+\tau_\downarrow)(1+\tau_\uparrow)}
< 1,\eeq and we have a geometric series that can be summed to
produce the transmission fraction for the downward hemisphere:
\beq T(\mu_+) = \frac{1 + \tau_\uparrow}{1 + \tau_\uparrow +
\tau_\downarrow}.\eeq A similar procedure for the upward
hemisphere can be carried out, arriving at the above sum plus
another factor of $R(\tau_\uparrow)$ accounting for the extra
reflection from the upper atmosphere required to make the flux
downward directed: \beqa T(\mu_-) &=&
\frac{\tau_\uparrow}{1+\tau_\uparrow}\frac{1}{1+\tau_\downarrow}
\nonumber\\ & & \times \sum_{p=0}^\infty \left[
\frac{\tau_\downarrow\tau_\uparrow}{(1+\tau_\downarrow)(1+\tau_\uparrow)}
\right]^p \nonumber\\ &=&
\frac{\tau_\uparrow}{1+\tau_\uparrow+\tau_\downarrow}.\eeqa

The total transmission of the source within the sandwich is \beqa
T&=& \frac{1}{2}\left[T(\mu_+)+T(\mu_-)\right] \nonumber\\ &=&
\frac{1/2+\tau_\uparrow}{1+\tau_\uparrow+\tau_\downarrow}.\eeqa In
the limit $\tau_\uparrow \gg \tau_\downarrow \gg 1$, we can see
that $T\rightarrow 1$, which allows us to define $\tau_\uparrow
\gg \tau_\downarrow$ as ``close to the ground,'' so no matter how
optically thick the atmosphere, auroral emission ``close to the
ground'' in a pure scattering atmosphere will reach the ground.
The limit $\tau_\uparrow \ll \tau_\downarrow$ is the Schuster
solution limit, in which $T\rightarrow 1/(2+2\tau_\downarrow)$.
Note that this limit is actually half of the Schuster transmission
because the source we consider is isotropic, while Schuster
defines the entire source flux to be incident on the atmosphere.

\bibliography{REFS}

\end{document}